\documentclass[final,5p,times,number]{elsarticle}

\usepackage{amssymb}
\journal{Measurement}  
\usepackage{booktabs}
\usepackage{array}
\usepackage{float}
\usepackage{textcomp}
\usepackage{tabularx}
\newcolumntype{L}[1]{>{\raggedright\arraybackslash\hsize=#1\hsize}X}
\usepackage{graphicx}
\usepackage{colortbl}
\usepackage{xcolor}
\usepackage{amsmath}
\usepackage{longtable}
\usepackage{tikz}
\usetikzlibrary{arrows.meta, positioning, calc}
\usepackage[normalem]{ulem} 

\usepackage{hyperref}
\hypersetup{colorlinks=true,linkcolor=black,citecolor=black,urlcolor=black}

\definecolor{addcolor}{RGB}{0,70,200}
\definecolor{delcolor}{RGB}{210,0,0}
\newcommand{\add}[1]{#1}                 
\newcommand{\del}[1]{}          
\newcommand{\delx}[1]{}                
\newenvironment{addblock}{}{}               
\newcommand{\yes}{\checkmark}
\newcommand{\no}{\texttimes}
\newcommand{\prt}{$\circ$}

\begin{document}

\begin{frontmatter}

\title{AquaSense-ICS: A Hybrid Real--Simulated Cyber-Physical Dataset and Leakage-Aware Benchmark for Smart Water Quality Monitoring and Attack Detection}

\author[crns]{Amal NEJI}
\author[crns]{Mohamed FOURATI}
\author[crns]{Lamia CHAARI FOURATI\corref{cor1}}
\ead{lamia.chaari@isims.usf.tn}
\cortext[cor1]{Corresponding author}

\address[crns]{Digital Research Center of Sfax (CRNS), Laboratory of Signals, Systems, Artificial Intelligence and Networks (SM@RTS), Sfax, Tunisia}

\begin{abstract}
Machine-learning research on water cyber-physical systems (CPS) is constrained by datasets that are either real but chemically poor, or simulated but free of sensing noise and attacks. We present \emph{AquaSense-ICS}, an open dataset and benchmark that co-registers (i) a 25.4\,h recording window (8.0\,h of recorded data, 26\% coverage) from a Raspberry Pi~5 edge node (TDS probe and BME680 ambient sensor, 1--2\,s sampling, 23{,}688 records), with (ii) a looped network of 13 nodes (10 junctions, 2 reservoirs, 1 tank) simulated in EPANET-MSX for six water-quality species, (iii) a parameterised library of cyber-attacks (false data injection, replay, denial of service, drift) and unintentional sensor faults with sample-level ground truth, and (iv) augmented and TimeGAN-generated sequences, flagged by an \texttt{origin} column. The attack, fault and water-quality labels are traceable to a seeded generator, a documented rule or a simulation input. We show that the most common way of benchmarking such hybrid data, random row splits and labels computed from the input features, yields near-perfect but uninformative scores (macro-F1 $=0.9998$ for Random Forest), and we replace it with a leakage-aware protocol using chronological splits and held-out attack types rather than label reconstruction. Under this protocol, no model exceeds the constant ``normal'' predictor for attack classification (macro-F1 $=0.465$), and the best detectors reach a mean AUROC of 0.64 on attack types unseen during training (0.35--0.93 depending on the type), leaving substantial headroom. Data and simulation files are released under CC BY 4.0 at \url{https://doi.org/10.5281/zenodo.23390541}, and attack generators and benchmark code under MIT.
\end{abstract}


\end{frontmatter}

\section{Introduction}\label{sec:intro}

\subsection{Motivation}
Water is a fundamental resource for human health, food security, industry, and ecosystem sustainability. However, urbanization, industrial discharge, agricultural pollution, and climate change are accelerating water stress worldwide.
{Around 2.4 billion people lived in water-stressed countries in 2020 \cite{UNSDG2023Goal6}, about two billion lack safely managed drinking water \cite{UNESCO2023WaterCrisis}, and microbiologically contaminated drinking water is estimated to cause about 505{,}000 diarrhoeal deaths each year \cite{WHO2023DrinkingWater},}
highlighting the critical importance of continuous water quality monitoring for public health and sustainability.

\subsection{IoT-Based Monitoring and Cyber-Physical Systems}
Traditional water quality monitoring relies on laboratory sampling, which is accurate but costly, delayed, and unsuitable for real-time decision-making. In contrast, IoT technologies enable continuous in-situ sensing and remote monitoring using low-cost sensors\add{~\cite{jan2021iotwater,camargo2023lowcost,Miller2023IoTWater}}. Parameters such as pH, conductivity, turbidity, temperature, and total dissolved solids can now be tracked in real time\add{~\cite{Zainurin2022WaterMonitoring}}. Moreover, modern water infrastructures increasingly rely on cyber-physical systems (CPS), which integrate sensing, communication, and control layers, improving efficiency while introducing new security risks.

\subsection{Artificial Intelligence for Smart Water Systems}
Artificial intelligence and machine learning have become essential for anomaly detection, forecasting, and intelligent water management. Deep learning models such as RNNs, Transformers, and graph neural networks have demonstrated strong performance in time-series analysis{, including attack detection in water distribution networks~\cite{Taormina2018DeepAE,tsiami2021cyberphysical}}. 
{Their progress, however, depends on datasets whose labels are trustworthy and whose evaluation protocols do not leak information from test to training data, a problem now recognised as a leading cause of irreproducible results in ML-based science~\cite{Kapoor2023Leakage}.}

\subsection{Cybersecurity Risks in Water Infrastructure}
The digitalization of water systems increases exposure to cyber-physical attacks. Water networks are high-value targets where malicious actions such as false data injection\add{~\cite{Liu2011FDI}}, replay attacks, denial-of-service, and drift attacks can silently manipulate sensor data and disrupt system operation\add{~\cite{Taormina2018BATADAL,Housh2018ModelBased}}. These threats highlight the need for datasets containing realistic and reproducible attack scenarios for robust anomaly detection research.

\subsection{Research Gap in Existing Datasets}
Despite recent advances, existing datasets remain limited. Benchmarks such as BattLeDIM {(on the L-TOWN network)~\cite{Vrachimis2022BattLeDIM}} and BATADAL{~\cite{Taormina2018BATADAL}} provide valuable simulation-based evaluation environments, but lack real IoT sensing and {(for BattLeDIM) attack scenarios}. {Testbed datasets such as SWaT and WADI~\cite{Mathur2016SWaT,Goh2016SWaTData,Ahmed2017WADI} contain real sensors and attacks but only a few water-quality analysers and no multi-species network chemistry.} Other datasets are often short-term, single-node, or parameter-limited, failing to capture the complexity of real water networks. To the best of our knowledge, no existing dataset simultaneously integrates real IoT sensing, EPANET-MSX multi-species simulation, cyber-physical attacks, time-series augmentation, and interpretable physicochemical labeling {, nor provides an evaluation protocol that controls for the leakage such hybrid data invite}.

\subsection{Proposed Dataset and Framework}
To address these limitations, this paper introduces a hybrid cyber-physical smart water quality dataset combining real IoT measurements, EPANET-MSX simulations, adversarial perturbations, and time-series augmentation within a unified temporal framework. The IoT layer is built using a Raspberry Pi 5 platform with BME680 and TDS sensors, capturing {ambient (air)} environmental and water quality variables over a 25.4,h monitoring window. The simulation layer uses EPANET-MSX\add{~\cite{Shang2008MSX,Shang2008MSXManual}} to model multi-species water dynamics including chlorine decay, pH variation, conductivity, dissolved oxygen, and trihalomethanes. A temporal fusion strategy aligns real and simulated data on a common elapsed-time axis. We state explicitly that the fusion is a \emph{co-registration}: the real sensor did not measure the simulated network, so cross-source relations in the fused table are synthetic by design and are documented as such.

\subsection{Research Questions}
The paper answers three questions:
\textbf{RQ1} Can a low-cost edge node and a multi-species network model be combined into a dataset whose every label is traceable and reproducible?
\textbf{RQ2} How much of the apparent performance on such hybrid data is caused by evaluation leakage, and what protocol removes it?
\textbf{RQ3} Under a leakage-aware protocol, how well do standard models detect and classify attacks, including attack types never seen in training, and which signals do they rely on?

\subsection{Main Contributions}
The main contributions are summarized as follows:
\begin{itemize}
    \item \textbf{Dataset:} An open hybrid dataset co-registering a real 25.4\,h edge-sensing recording and a 6-species EPANET-MSX network simulation, released as linked raw, simulated, attacked and fused tables with a data dictionary (Section~\ref{sec:availability}).
    \item \textbf{Attack and fault library:} A seeded generator for five attack types (including stealthy FDI) and three fault types with sample-level labels, applied to real and simulated channels (Section~\ref{sec:perturb}).
    \item \textbf{Traceable labelling:} Separate water-quality, attack and fault labels, each derived from regulatory thresholds or from the generator (Section~\ref{sec:fusion}).
    \item \textbf{Leakage-aware benchmark:} We quantify how random splits and feature-derived labels inflate scores, and define tasks with chronological and leave-attack-type-out splits (Sections~\ref{sec:benchmark}--\ref{sec:xai}).
    \item \textbf{Reproducibility:} Code, the \texttt{.inp}/\texttt{.msx} files, random seeds and a datasheet~\cite{Gebru2021Datasheets} following FAIR principles~\cite{Wilkinson2016FAIR}.
\end{itemize}

\subsection{Paper Organization}
\begin{figure*}[t]
\centering

\begin{tikzpicture}[
    node distance=0.75cm and 0.65cm,
    box/.style={
        rounded corners=5pt,
        minimum width=4.0cm,
        minimum height=1.05cm,
        align=center,
        font=\small\bfseries,
        draw=black!55,
        line width=0.8pt
    },
    intro/.style={box, fill=purple!18},
    related/.style={box, fill=green!18},
    sensing/.style={box, fill=orange!25},
    simulation/.style={box, fill=red!18},
    perturb/.style={box, fill=lime!25},
    fusion/.style={box, fill=gray!17},
    analysis/.style={box, fill=blue!18},
    benchmark/.style={box, fill=cyan!18},
    xai/.style={box, fill=violet!16},
    availability/.style={box, fill=orange!18},
    conclusion/.style={box, fill=cyan!25},
    arrow/.style={
        -{Latex[length=2.2mm]},
        line width=0.9pt,
        draw=red!60!black
    }
]

\node[intro] (intro) {Introduction};

\node[related, right=of intro] (related)
{Related Datasets};

\node[sensing, right=of related] (acq)
{Real Data Acquisition};

\node[analysis, below=of intro] (analysis)
{Dataset Analysis};

\node[perturb, right=of analysis] (perturb)
{Augmentation, Attacks\\and Faults};

\node[simulation, right=of perturb] (sim)
{EPANET--MSX\\Simulation};

\node[fusion, below=of analysis] (fusion)
{Temporal Fusion\\and Labelling};

\node[benchmark, right=of fusion] (benchmark)
{Benchmark Evaluation};

\node[xai, right=of benchmark] (xai)
{Explainability\\Analysis};

\node[availability, below=of xai] (availability)
{Dataset Availability\\and Release};

\node[conclusion, left=of availability] (conclusion)
{Conclusion, Limitations\\and Future Work};

\draw[arrow] (intro) -- (related);
\draw[arrow] (related) -- (acq);

\draw[arrow]
(acq.east)
to[out=0,in=0,looseness=1.35]
(sim.east);

\draw[arrow] (sim) -- (perturb);
\draw[arrow] (perturb) -- (analysis);

\draw[arrow]
(analysis.west)
to[out=180,in=180,looseness=1.35]
(fusion.west);

\draw[arrow] (fusion) -- (benchmark);
\draw[arrow] (benchmark) -- (xai);

\draw[arrow]
(xai.east)
to[out=0,in=0,looseness=1.3]
(availability.east);

\draw[arrow] (availability) -- (conclusion);

\end{tikzpicture}

\caption{Roadmap of the paper and methodological workflow. The study
progresses from related datasets and real-world data acquisition to
EPANET--MSX simulation, cyber-physical perturbation generation, temporal
fusion and labelling, dataset analysis, benchmarking, explainability,
and finally dataset release and discussion of limitations and future work.}
\label{fig:roadmap}
\end{figure*}
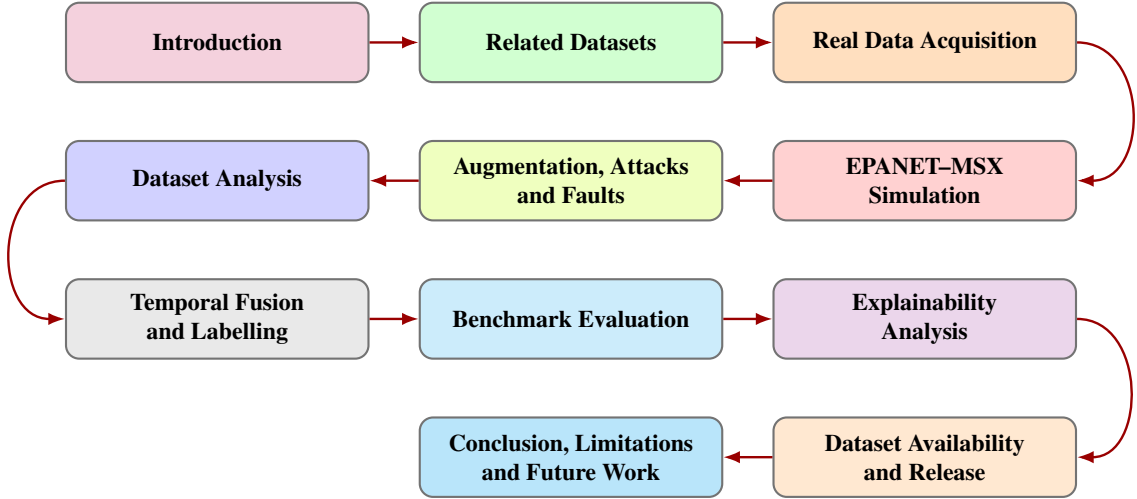

The remainder of this paper is organized as follows. {Section~\ref{sec:related} reviews related datasets. Section~\ref{sec:acq} describes real data acquisition, Section~\ref{sec:sim} the EPANET-MSX simulation, and Section~\ref{sec:perturb} augmentation, attacks and faults. Section~\ref{sec:fusion} covers fusion and labelling, Section~\ref{sec:eda} dataset analysis, Section~\ref{sec:benchmark} the benchmark and Section ~\ref{sec:xai} explainability.
Section~\ref{sec:availability} documents the release and Section~\ref{sec:conclusion} concludes with limitations.}
Figure~\ref{fig:roadmap} illustrates the overall organization and methodological workflow of the proposed framework.

\subsection{List of Acronyms}
The acronyms used in this paper are summarized in Table~\ref{tab:acronyms}.

\begin{table*}[!ht]
\centering
\small
\renewcommand{\arraystretch}{1.1}
\setlength{\tabcolsep}{4pt}
\caption{{Principal} acronyms used in this paper}
\label{tab:acronyms}
\begin{tabular}{@{} p{2.2cm} p{5.8cm} p{2.2cm} p{5.8cm} @{}}
\toprule
\textbf{Acronym} & \textbf{Definition} & \textbf{Acronym} & \textbf{Definition} \\
\midrule
ADC & Analog-to-digital converter & MSX & Multi-Species eXtension (EPANET) \\
BATADAL & Battle of the Attack Detection Algorithms & NTU & Nephelometric turbidity unit \\
BattLeDIM & Battle of the Leakage Detection and Isolation Methods & RH & Relative humidity \\
ClD & Wall chlorine demand (MSX wall species) & SHAP & SHapley Additive exPlanations \\
COND & Electrical conductivity & SWaT & Secure Water Treatment testbed \\
CPS & Cyber-physical system & TDS & Total dissolved solids \\
DO & Dissolved oxygen & THM & Trihalomethanes \\
DoS & Denial of service & TSTR & Train on synthetic, test on real \\
FDI & False data injection & TURB & Turbidity \\
FAR & False-alarm rate & VOC & Volatile organic compounds \\
LOAO & Leave-one-attack-type-out split & WADI & Water Distribution testbed \\
LOEO & Leave-one-event-type-out split & WDN & Water distribution network \\
\bottomrule
\end{tabular}
\end{table*}

\section{Related Work}\label{sec:related}
\subsection{Overview of Existing Datasets for Smart Water Systems}
The development of data-driven methods for smart water management has been supported by a growing but still limited number of publicly available datasets. These resources can be broadly categorized into four classes: real IoT sensing datasets, hydraulic simulation benchmarks, cyber-physical attack datasets, and hybrid resources. Table~\ref{tab:related_work_table} provides a comparative summary of the most representative existing datasets alongside the proposed benchmark.

\subsection{\texorpdfstring{Real {Testbed and Sensing Datasets}}{Real Testbed and Sensing Datasets}}
 The Secure Water Treatment (SWaT) {testbed}, introduced by Mathur and Tippenhauer (2016){~\cite{Mathur2016SWaT}}, {is} a fully operational scaled-down water treatment facility{; its public dataset~\cite{Goh2016SWaTData}} includes both normal operations and 36 {executed} cyberattack scenarios recorded over 11 days{\cite{Mathur2016SWaT}}. Although SWaT provides real sensor measurements and attack labels, it targets a treatment plant rather than a distribution network, and does not include hydraulic simulation, multi-species chemical modeling, or time-series augmentation. The WADI (Water Distribution) dataset, proposed by Ahmed et al. (2017) as a follow-up to SWaT, extends the testbed to a water distribution infrastructure and similarly provides labeled attack scenarios.\cite{Ahmed2017WADI} However, both SWaT and WADI remain confined to controlled laboratory environments and {include only a few water-quality analysers (e.g., pH, conductivity and ORP in SWaT); network-scale species such as chlorine residual, dissolved oxygen and 
 
 trihalomethanes are absent, and access requires a request to the operating institution.}

The WaterBench dataset, released by the Open Water AI Initiative (2022), provides real IoT measurements from deployed sensors but lacks attack scenarios, quality labels, and augmentation pipelines, limiting its applicability for cybersecurity and machine learning benchmarking.\cite{WaterBench2023} More broadly, survey papers covering IoT water quality datasets (2019--2024) consistently highlight the absence of unified resources that combine sensing, simulation, security, and labeling within a single open benchmark.\cite{EPANETStudy2019}

{Reviews of low-cost and IoT water-quality monitoring~\cite{camargo2023lowcost,Miller2023IoTWater,jan2021iotwater} report that most deployments are short, single-site and unlabelled, and that openly released raw data with documented calibration remain rare.}

\subsection{Hydraulic Simulation Benchmarks}
Simulation-based datasets have been the dominant resource in water network research. EPANET, developed by Rossman (2000) at the US Environmental Protection Agency{~\cite{EPANET2000,Rossman2020EPANET22}}, remains the standard tool for hydraulic modeling of water distribution systems and has been widely used to generate synthetic datasets for pressure monitoring, leakage detection, and demand forecasting. However, standard EPANET {tracks a single water-quality constituent and does} not capture the multi-species chemical dynamics including chlorine decay, trihalomethane formation, pH variation, and dissolved oxygen evolution that are critical for water quality analysis. The EPANET-MSX extension\add{~\cite{Shang2008MSX}} addresses this limitation by enabling multi-species reaction modeling within distribution networks, yet datasets generated using EPANET-MSX remain scarce and have not been integrated with real sensor observations or cybersecurity scenarios. The BattLeDIM competition (Vrachimis et al., 2022), based on the L-TOWN network, provided a valuable benchmark for leakage detection and isolation using SCADA-simulated measurements of flow, pressure, and tank levels. While BattLeDIM established a rigorous evaluation framework, it relies entirely on simulated data and does not incorporate water quality parameters{ or cyber-attacks}.\cite{Vrachimis2022BattLeDIM} The LeakDB benchmark similarly offers simulated leakage scenarios across multiple network topologies but lacks quality modeling and real sensing integration.\cite{Vrachimis2018LeakDB}

\subsection{Cyber-Physical Attack Datasets}
The BATADAL (Battle of the Attack Detection Algorithms) competition (Taormina et al., 2018) represents the most widely cited benchmark for cyber-physical attack detection in water distribution systems. BATADAL provided simulated SCADA observations from the C-Town network under {a set of labelled attack events}, enabling the evaluation of detection algorithms across multiple teams. Despite its significant impact, BATADAL relies exclusively on hydraulic simulation and does not include real IoT measurements, water quality parameters, or time-series augmentation.{\cite{Taormina2018BATADAL}} {It spawned a line of detectors based on autoencoders~\cite{Taormina2018DeepAE}, hydraulic models~\cite{Housh2018ModelBased}, classical anomaly detection~\cite{Abokifa2019ML} and temporal graph networks~\cite{tsiami2021cyberphysical}, all evaluated on hydraulic (not chemical) signals.} The KIOS research center datasets associated with BattLeDIM provide structured anomaly labels but similarly lack cybersecurity enrichment and multi-species chemical variables\cite{BattLeDIM2020}.

\subsection{Limitations of Existing Resources and Motivation}
Despite the progress outlined above, a critical gap persists in the literature. As summarized in Table~\ref{tab:related_work_table}, no existing publicly available dataset simultaneously {offers} real {edge} sensing, multi-species hydraulic and chemical simulation, {parameterised attack and fault labels}, time-series augmentation {kept separate from real data}, interpretable physicochemical quality labeling, and full reproducibility. {Equally important, none of these resources specifies an evaluation protocol that guards against temporal and label leakage~\cite{Bergmeir2012CV,Kapoor2023Leakage}.} This gap motivates the design of the proposed {dataset and benchmark}.

\begin{table*}[t]
\centering
\footnotesize
\setlength{\tabcolsep}{3.5pt}
\renewcommand{\arraystretch}{1.2}
\caption{Comparative analysis of existing smart water datasets {(\yes\ = yes, \prt\ = partial, \no\ = no). Columns revised to separate real water-quality sensing from ambient sensing and to add the evaluation protocol.}}
\label{tab:related_work_table}
\begin{tabular}{lccccccccl}
\toprule
\textbf{Dataset} & \textbf{Real} & \textbf{Real WQ} & \textbf{Network} & \textbf{Multi-} & \textbf{Attack} & \textbf{Fault} & \textbf{Synthetic /} & \textbf{Leakage-} & \textbf{Access} \\
 & \textbf{sensors} & \textbf{analysers} & \textbf{simulation} & \textbf{species} & \textbf{labels} & \textbf{labels} & \textbf{augmented} & \textbf{aware split} & \\
\midrule
SWaT \cite{Mathur2016SWaT,Goh2016SWaTData} & \yes & \prt & \no & \no & \yes & \no & \no & \no & On request \\
WADI \cite{Ahmed2017WADI} & \yes & \prt & \no & \no & \yes & \no & \no & \no & On request \\
BATADAL \cite{Taormina2018BATADAL} & \no & \no & \yes & \no & \yes & \no & \no & \prt & Open \\
BattLeDIM / L-TOWN \cite{Vrachimis2022BattLeDIM} & \no & \no & \yes & \no & \no & \no & \no & \prt & Open \\
LeakDB \cite{Vrachimis2018LeakDB} & \no & \no & \yes & \no & \no & \no & \no & \no & Open \\
\midrule
\textbf{AquaSense-ICS (this work)} & \yes & \prt & \yes & \yes & \yes & \yes & \yes & \yes & Open (CC BY 4.0) \\
\bottomrule
\end{tabular}
\end{table*}

\section{Measurement Architecture and Real Data Acquisition}\label{sec:acq}

This section presents the experimental platform developed for real-time monitoring of water quality and environmental conditions. It describes the proposed IoT architecture, sensing devices, hardware/software configuration, and the methodology adopted for data acquisition and preprocessing. The platform was designed to serve as a cyber-physical testbed, enabling both machine learning model training and anomaly detection evaluation under realistic and controlled operating conditions.

\subsection{Proposed IoT Architecture}

\begin{figure*}[t]
\centering
\includegraphics[width=0.9\textwidth]{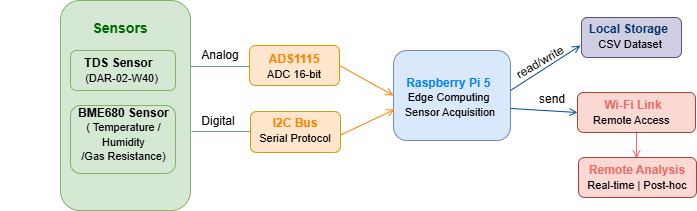}
\caption{Overall IoT architecture and data flow of the proposed monitoring platform.}
\label{fig:iot_architecture}
\end{figure*}

The proposed monitoring platform is centered on a Raspberry Pi 5, which serves as the main edge-computing unit responsible for sensor interfacing, data acquisition, local storage, and wireless transmission. The system integrates water-quality sensing and environmental monitoring modules within a unified IoT architecture, as illustrated in Figure~\ref{fig:iot_architecture}.

The Raspberry Pi communicates with digital sensors through the I2C serial protocol and acquires analog measurements via an external ADS1115 analog-to-digital converter (ADC). Collected measurements are stored locally in CSV format and can be remotely accessed via Wi-Fi for real-time supervision and post-hoc analysis. The edge-first design minimizes latency and network dependency, making the system suitable for deployment in resource-constrained environments.

\subsection{Sensors and Hardware Configuration}

Two main sensing modules were employed in the experimental setup, selected for their complementary measurement capabilities, compact form factors, and compatibility with the Raspberry Pi GPIO interface:

\begin{itemize}
    \item \textbf{TDS Sensor (DAR-02-W40):}
    \begin{itemize}
        \item Measures Total Dissolved Solids (TDS) in water, reflecting overall salinity and impurity concentration.
        \item Outputs an analog voltage signal proportional to ion concentration, which is digitized by the ADS1115 ADC before transmission to the Raspberry Pi.
        \item Selected for its cost-effectiveness and suitability for embedded analog acquisition pipelines.
    \end{itemize}

    \item \textbf{BME680 Environmental Sensor (Bosch):}
    \begin{itemize}
        \item A multifunction MEMS sensor providing simultaneous measurement of \add{air} temperature ($^{\circ}$C), relative humidity (\%), barometric pressure (hPa), and gas resistance ($\Omega$)\add{~\cite{BME680Datasheet}}.
        \item Gas resistance values are inversely correlated with volatile organic compound (VOC) concentration, making the sensor sensitive to gaseous pollution events.
        \item Communicates directly with the Raspberry Pi via the I2C bus using Adafruit's CircuitPython library.
        \item { The BME680 sensor was positioned above the water surface to characterise the \emph{ambient} environment of the monitoring station. During perturbation experiments, the sensor was progressively moved closer to and then farther away from the perturbation source to capture the effect of varying proximity. Barometric pressure is measured by the BME680 but is not consistently recorded across all experimental runs and is excluded from the release.}

    \end{itemize}
\end{itemize}

The physical testbed consists of a water container simulating a monitored water source. External perturbation sources are introduced in the vicinity of the sensing unit to emulate realistic environmental disturbances.

\subsection{TDS Measurement and Limitations}\label{sec:calib}
The TDS probe provides an analog voltage $V$ digitised by the ADS1115. The released column \texttt{tds\_value} is the output of the acquisition script,
\begin{equation}
    \mathrm{TDS} = \frac{V \times 0.5}{1 + \alpha (T_a - 25)},\qquad \alpha = 0.02\,^{\circ}\mathrm{C}^{-1},
    \label{eq:tds_conversion}
\end{equation}
where $T_a$ is the BME680 air temperature. This expression is a scaled voltage, not a calibrated concentration: the probe was not calibrated against standard solutions, the raw voltage was not retained, and the values (0.007--1.196) are not expressed in ppm. The column is therefore released without a unit. No immersed temperature probe was available, so the compensation uses the ambient air temperature, which adds a further uncertainty.

\subsection{Measurement Ranges and Operating Thresholds}

To ensure reliable interpretation of sensor readings and to support rule-based anomaly labeling, operating thresholds were defined based on WHO drinking water quality guidelines \cite{WHO2022} and standard {indoor-environment} reference ranges.

Table~\ref{tab:thresholds} summarizes the operational ranges used in this study. \add{Air} temperature was considered normal within the range of 10--30$^{\circ}$C, while values outside this interval were flagged as anomalous. Relative humidity was constrained between 25--80\%, corresponding to stable indoor environmental conditions. Gas resistance values from the BME680 sensor were interpreted as indicators of air quality, with acceptable operation between 8,000 and 200,000 $\Omega$, where lower values indicate potential gas pollution. For water quality, Total Dissolved Solids (TDS) values below 600 ppm were considered acceptable, with further categorization into excellent quality ($<$300 ppm), {good} (300--600 ppm), and degraded conditions ($>$600 ppm), according to {the WHO palatability ranges; WHO sets no health-based guideline value for TDS} {\cite{WHO2022}}.

\begin{table}[H]
\centering
\caption{Measurement ranges and operating thresholds.}
\label{tab:thresholds}
\resizebox{\columnwidth}{!}{%
\begin{tabular}{lllll}
\toprule
\textbf{Parameter} & \textbf{Unit} & \textbf{Normal Range} & \textbf{Threshold / Classification} & \textbf{Ref.} \\
\midrule
\add{Air} temperature      & $^{\circ}$C & 10-30         & $<$10 or $>$30: abnormal                        & \add{\cite{ASHRAE55}} \\
Relative Humidity & \%          & 25-80         & $<$25 or $>$80: abnormal                        & \cite{ASHRAE55} \\
Gas Resistance   & $\Omega$    & 8k-200k       & $<$8,000: poor air quality                      & \cite{BME680Datasheet} \\
TDS              & ppm         & $<$600         & $<$300: excellent; 300-600: good; $>$600: poor & \cite{WHO2022} \\
\bottomrule
\end{tabular}%
}
\end{table}
These thresholds were subsequently used for automated anomaly detection and binary {ambient-anomaly labelling (\texttt{env\_flag}), which is distinct from the water-quality label of Section~\ref{sec:labels}}.

\subsection{Software Environment}

The following software stack was used to operate the acquisition system, process data, and conduct downstream analysis:

\begin{itemize}
\item Python 3.11\add{:} real-time sensor reading, data logging, and preprocessing scripts.
\item Adafruit CircuitPython BME680 library: I2C communication and sensor configuration.
\item ADS1x15 Python library: ADC interfacing for TDS analog signal acquisition.
\item Local CSV storage: timestamped records saved to the Raspberry Pi filesystem {, with timestamps in UTC synchronised by NTP}.
\item RealVNC Viewer: secure remote desktop access to the Raspberry Pi for monitoring and control.
\item Google Colab (Python): cloud-based environment for data analysis, feature engineering, and ML model training, using Pandas, NumPy, Matplotlib, Scikit-learn, and TensorFlow{; 
exact package versions are pinned in 

\texttt{requirements.txt} in the followin repository \url{https://github.com/nejiamal3-oss/aquasense}}.
\end{itemize}

\subsection{Data Acquisition Methodology}

Data collection spanned a 25.4-hour window (from 19 May 2025 at 12:39:08 to 20 May 2025 at 14:01:00), of which 8.0\,h contain recorded samples (26\% coverage). The record is interrupted by 23 gaps longer than 60\,s, the longest lasting about 10.1\,h. Within recorded periods, consecutive samples are most often separated by 1--2\,s (median step 1\,s); the mean step of 3.9\,s is inflated by the gaps.This acquisition rate was chosen to enable fine-grained temporal monitoring of sensor dynamics, particularly transient pollution events, while maintaining manageable storage overhead. Each recorded sample is timestamped and contains four measurement fields: TDS, temperature, relative humidity, and gas resistance. The main acquisition statistics are summarized in Table~\ref{tab:acquisition}.

\begin{table}[H]
\centering
\caption{Summary of the acquisition campaign.}
\label{tab:acquisition}
\resizebox{\columnwidth}{!}{%
\begin{tabular}{ll}
\toprule
\textbf{Metric} & \textbf{Value} \\
\midrule
Total acquisition window & 25.4 hours (8.0\,h recorded) \\
Sampling interval & 1--2 seconds (median 1\,s) \\
Number of records & 23{,}688 samples \\
Missing samples (vs.\ a nominal 1\,s step) & 67{,}625 (74.1\%) \\
Raw dataset size & $\sim$2 MB (CSV format) \\
Measured parameters & TDS, Temp., Humidity, Gas Resistance \\
\bottomrule
\end{tabular}%
}
\end{table}

\textit{Note:} The 25.4-hour window (8.0,h of recorded data) is used as an initial proof-of-concept to validate the proposed cyber-physical IoT pipeline. The same deployment framework is  scalable to long-term monitoring (weeks to months), which is expected to enhance temporal generalization and improve the robustness of learned models under diverse environmental conditions.

\subsection{Experimental Perturbation Scenarios}

To evaluate system responsiveness under dynamic and adverse conditions, three categories of controlled perturbations were introduced during the acquisition period. These disturbances were designed to emulate realistic pollution and environmental stress scenarios encountered in indoor or semi-controlled water monitoring deployments:

\begin{itemize}
    \item \textbf{Cigarette smoke (VOC/gaseous pollution):}
    Simulates indoor air quality degradation due to volatile organic compounds (VOCs), resulting in sharp decreases in gas resistance readings.

    \item \textbf{Cup of water (humidity spike):}
    A water-filled container placed near the sensor to simulate evaporation-induced humidity increases, emulating condensation-prone environments.

    \item \textbf{Candle flame (thermal anomaly):}
    Introduced as a localized heat source to emulate thermal anomalies such as those caused by nearby equipment or solar radiation.
\end{itemize}

The impact of each disturbance is reflected in the corresponding sensor channels. However, explicit event start and end times are not available in the dataset. These disturbances affect the ambient environment rather than the water itself. Consequently, they are not used as ground-truth event labels for a dedicated event-detection task.

\subsection{Data Preprocessing Pipeline}

The released dataset keeps every raw sample in physical units together with quality flags; the steps below produce the analysis-ready version and are implemented as a script so that users may re-run or skip them.

\begin{itemize}

\item \textbf{Quality control:}
The raw record contains no physically impossible reading (maximum humidity 73.6\%, no negative gas resistance or TDS value). The fused file keeps the \emph{observed} channels after fault and attack injection, whose extreme values are identified by \texttt{fault\_type}.

\item \textbf{Missing values:}
The four sensor channels have no missing value, so no imputation is applied.

\item \textbf{Feature normalization:}
Normalisation is \emph{not} applied to the released data. In the benchmark it is fitted on the training split only; gas resistance is log-transformed before z-scoring because of its range of more than two orders of magnitude.

\item \textbf{Temporal windowing:}
The benchmark uses windows of $W=30$ samples with stride 5; a window is labelled by its last sample.

\item \textbf{Label generation:}
See Section~\ref{sec:labels}.

\end{itemize}

\section{Hydraulic and Chemical Simulation using EPANET-MSX}\label{sec:sim}

This section presents the simulation framework used to generate synthetic hydraulic and water quality data for the cyber-physical dataset. The objective is to complement real IoT sensor measurements with physically consistent, multi-species hydraulic modeling, thereby enabling {a benchmark that combines} both real-world variability and controlled, reproducible simulation scenarios.

\subsection{EPANET-MSX Simulation Framework}

\textbf{EPANET-MSX} (Multi-Species eXtension) is an open-source extension of the U.S.\ EPA EPANET hydraulic solver that enables multi-species reactive transport modeling in pressurized water distribution networks (WDNs)\add{~\cite{Shang2008MSX,Shang2008MSXManual}}. Unlike standard EPANET, which supports only single-species {water-quality tracking}, EPANET-MSX allows the simultaneous representation of coupled chemical reactions involving multiple water quality species in both the \textit{bulk water} and \textit{pipe-wall (biofilm)} domains.

The simulation framework integrates hydraulic dynamics with multi-species chemical evolution, capturing the following interacting phenomena:

\begin{itemize}
    \item \textbf{Chlorine (Cl) decay}\add{:} first-order bulk degradation coupled with biofilm wall demand.
    \item \textbf{pH variation}\add{:} dynamic equilibration driven by disinfection chemistry.
    \item \textbf{Turbidity (TURB)}\add{:} evolution of suspended particle load and water clarity.
    \item \textbf{Dissolved Oxygen (DO)}\add{:} consumption through oxidation reactions coupled with chlorine dynamics.
    \item \textbf{Conductivity (COND)}\add{:} variation in ionic concentration along the distribution network.
    \item \textbf{Trihalomethane formation (THM)}\add{:} disinfection by-products (DBPs) formed as a chlorine-dependent reaction pathway, regulated under EPA/WHO drinking water standards.
    \item \textbf{Wall Chlorine Demand (ClD)}\add{:} biofilm-induced chlorine consumption on internal pipe surfaces.
\end{itemize}

The MSX configuration file defines the reaction kinetics governing each species in both bulk and wall phases, using coupled ordinary differential equations (ODEs) solved simultaneously at each time step across all network nodes and pipe segments.

\subsection{Hydraulic Network Topology}

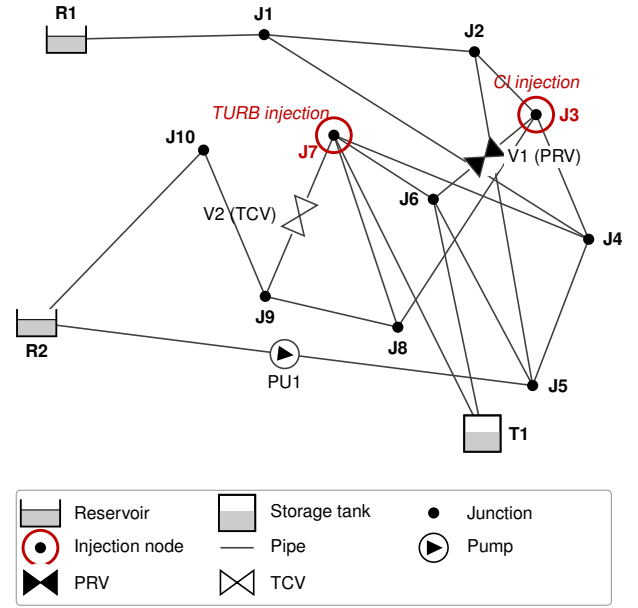
\begin{figure}[H]
\centering

\begin{tikzpicture}[
    x=1cm, y=1cm,
    font=\footnotesize\sffamily,
    pipe/.style      = {line width=0.6pt, draw=black!75},
    junction/.style  = {circle, fill=black, inner sep=0pt, minimum size=4.2pt},
    lbl/.style       = {font=\scriptsize\sffamily\bfseries, inner sep=1.2pt},
    linklbl/.style   = {font=\scriptsize\sffamily, inner sep=1pt},
    annot/.style     = {font=\scriptsize\sffamily\itshape, red!75!black, inner sep=1pt},
    reservoir/.pic = {
        \fill[white] (-2.6mm,-1.8mm) rectangle (2.6mm,1.8mm);
        \fill[black!20] (-2.6mm,-1.8mm) rectangle (2.6mm,0.4mm);
        \draw[line width=0.7pt] (-2.6mm,1.8mm) -- (-2.6mm,-1.8mm) -- (2.6mm,-1.8mm) -- (2.6mm,1.8mm);
        \draw[line width=0.4pt] (-2.6mm,0.4mm) -- (2.6mm,0.4mm);
    },
    tank/.pic = {
        \filldraw[fill=white, line width=0.7pt] (-2.4mm,-2.4mm) rectangle (2.4mm,2.4mm);
        \fill[black!20] (-2.4mm,-2.4mm) rectangle (2.4mm,0.2mm);
        \draw[line width=0.7pt] (-2.4mm,-2.4mm) rectangle (2.4mm,2.4mm);
    },
    pump/.pic = {
        \filldraw[fill=white, line width=0.6pt] (0,0) circle (1.9mm);
        \fill (-0.9mm,-1.0mm) -- (1.2mm,0) -- (-0.9mm,1.0mm) -- cycle;
    },
    prv/.pic = {
        \fill[white] (0,0) circle (2.6mm);
        \filldraw[fill=black, line width=0.5pt] (-2.2mm,-1.6mm) -- (2.2mm,1.6mm) -- (2.2mm,-1.6mm) -- (-2.2mm,1.6mm) -- cycle;
    },
    tcv/.pic = {
        \fill[white] (0,0) circle (2.6mm);
        \filldraw[fill=white, line width=0.6pt] (-2.2mm,-1.6mm) -- (2.2mm,1.6mm) -- (2.2mm,-1.6mm) -- (-2.2mm,1.6mm) -- cycle;
    },
    injring/.pic = { \draw[red!75!black, line width=1.1pt] (0,0) circle (2.3mm); },
]
 
\coordinate (J1) at (3.008,5.275);
\coordinate (J2) at (5.789,5.048);
\coordinate (J3) at (6.605,4.217);
\coordinate (J4) at (7.300,2.569);
\coordinate (J5) at (6.559,0.635);
\coordinate (J6) at (5.245,3.098);
\coordinate (J7) at (3.930,3.945);
\coordinate (J8) at (4.776,1.406);
\coordinate (J9) at (3.023,1.814);
\coordinate (J10) at (2.207,3.748);
\coordinate (R1) at (0.393,5.214);
\coordinate (R2) at (0.000,1.466);
\coordinate (T1) at (5.894,0.000);
 
\draw[pipe] (R1) -- (J1); 
\draw[pipe] (J1) -- (J2); 
\draw[pipe] (J2) -- (J3); 
\draw[pipe] (J3) -- (J4); 
\draw[pipe] (J4) -- (J5); 
\draw[pipe] (J5) -- (J6); 
\draw[pipe] (J6) -- (J7); 
\draw[pipe] (J7) -- (J8); 
\draw[pipe] (J8) -- (J9); 
\draw[pipe] (J9) -- (J10); 
\draw[pipe] (J10) -- (R2); 
\draw[pipe] (J2) -- (J5); 
\draw[pipe] (J4) -- (J7); 
\draw[pipe] (J3) -- (J8); 
\draw[pipe] (J6) -- (T1); 
\draw[pipe] (T1) -- (J7); 
\draw[pipe] (J1) -- (J4); 
 
\draw[pipe] (R2) -- pic[pos=0.5, sloped]{pump} coordinate[pos=0.5](PU1m) (J5);
\node[linklbl, below=2.4mm] at (PU1m) {PU1};
\draw[pipe] (J3) -- pic[pos=0.5, sloped]{prv} coordinate[pos=0.5](V1m) (J6);
\node[linklbl, right=2.6mm, fill=white] at (V1m) {V1 (PRV)};
\draw[pipe] (J7) -- pic[pos=0.5, sloped]{tcv} coordinate[pos=0.5](V2m) (J9);
\node[linklbl, left=2.6mm] at (V2m) {V2 (TCV)};
 
\pic at (R1) {reservoir};
\pic at (R2) {reservoir};
\pic at (T1) {tank};
\node[junction] at (J1) {};
\node[junction] at (J2) {};
\node[junction] at (J3) {};
\node[junction] at (J4) {};
\node[junction] at (J5) {};
\node[junction] at (J6) {};
\node[junction] at (J7) {};
\node[junction] at (J8) {};
\node[junction] at (J9) {};
\node[junction] at (J10) {};
\pic at (J3) {injring};
\pic at (J7) {injring};
\node[lbl, above=2.2mm] at (R1) {R1};
\node[lbl, below=2.4mm] at (R2) {R2};
\node[lbl, right=3.0mm] at (T1) {T1};
\node[lbl, above=1.4mm] at (J1) {J1};
\node[lbl, above=1.4mm] at (J2) {J2};
\node[lbl, right=2.6mm, red!75!black] at (J3) {J3};
\node[lbl, right=1.4mm] at (J4) {J4};
\node[lbl, right=1.6mm] at (J5) {J5};
\node[lbl, left=1.4mm] at (J6) {J6};
\node[lbl, below left=0.8mm and 1.6mm, red!75!black] at (J7) {J7};
\node[lbl, below=1.4mm] at (J8) {J8};
\node[lbl, below=1.4mm] at (J9) {J9};
\node[lbl, above left=0.4mm and 0.6mm] at (J10) {J10};
\node[annot, above=2.6mm] at (J3) {Cl injection};
\node[annot, above left=1.2mm and 0.2mm] at (J7) {TURB injection};
 
\begin{scope}[shift={(0.05,-1.05)}, font=\scriptsize\sffamily]
  \def\cA{0}\def\cB{2.6}\def\cC{5.2}
  \draw[black!40, line width=0.4pt, rounded corners=1.5pt] (-0.32,-1.22) rectangle (7.55,0.30);
  \pic at (\cA,0) {reservoir};   \node[anchor=west] at (\cA+0.32,0) {Reservoir};
  \pic at (\cB,0) {tank};        \node[anchor=west] at (\cB+0.32,0) {Storage tank};
  \node[junction] at (\cC,0) {}; \node[anchor=west] at (\cC+0.32,0) {Junction};
  \node[junction] at (\cA,-0.46) {}; \pic at (\cA,-0.46) {injring};
                                 \node[anchor=west] at (\cA+0.32,-0.46) {Injection node};
  \draw[pipe] (\cB-0.22,-0.46) -- (\cB+0.22,-0.46);
                                 \node[anchor=west] at (\cB+0.32,-0.46) {Pipe};
  \pic at (\cC,-0.46) {pump};    \node[anchor=west] at (\cC+0.32,-0.46) {Pump};
  \pic at (\cA,-0.92) {prv};     \node[anchor=west] at (\cA+0.32,-0.92) {PRV};
  \pic at (\cB,-0.92) {tcv};     \node[anchor=west] at (\cB+0.32,-0.92) {TCV};
\end{scope}
 
\end{tikzpicture}
\caption{Hydraulic network topology generated using EPANET.}
\label{fig:network}
\end{figure}

The simulated network (Fig.~\ref{fig:network}) represents a single pressure
zone of a municipal distribution system, comparable to a district metered area
(DMA). We model one zone rather than a full city network because utilities
operate, monitor, and isolate their systems at the zone level. Sensor
placement, anomaly detection, and incident response are therefore normally
carried out at this scale. The network has 16.8\,km of pipes with diameters of
250--300\,mm and a total base demand of 59.5\,L/s (Table~\ref{tab:network}).

The topology was designed to meet three requirements. \textit{(i) Coverage of
the attack surface:} it includes every class of component whose manipulation
can compromise water quality or supply, namely two fixed-head sources (R1, R2)
with different source quality, a storage tank (T1), a pump (PU1), a
pressure-reducing valve (PRV-1), and a throttle control valve (TCV-1).
\textit{(ii) Hydraulic diversity:} looped sections produce redundant paths and
flow reversals, while branched sections produce long residence times. \textit{(iii) Interpretability:} with 13 nodes, each detected
anomaly can be traced to a physical propagation path, so the SHAP attributions
in Section~\ref{sec:xai} can be checked against known hydraulic cause and effect.

Networks of this scale remain standard in recent water-security and
water-quality studies: EPANET Net1 (9 junctions) is the main case study in
recent work on stealthy cyberattacks against network hydraulics~\cite{albustami2025breaking}
and on multi-species water-quality modelling and control validated against
EPANET-MSX~\cite{elsherif2023control,elsherif2024comprehensive}, while the Hanoi network (32 nodes) is used for machine-learning contamination detection~\cite{li2025online}. Our network
is of comparable size but contains more control elements (a pump, a PRV and a
TCV, two sources and a tank).
The network (\texttt{.inp}) and reaction (\texttt{.msx}) files are released so
the same pipeline can be applied to larger systems.

\begin{table}[H]
\centering
\caption{Hydraulic network components}
\label{tab:network}
\resizebox{\columnwidth}{!}{%
\begin{tabular}{llcl}
\hline
\textbf{Component} & \textbf{ID(s)} & \textbf{Count} & \textbf{Role} \\
\hline
Reservoirs      & R1, R2           & 2  & Fixed-head water sources \\
Junction nodes  & J1--J10          & 10 & Demand / sampling points \\
Storage tank    & T1               & 1  & Buffer reservoir \\
Pipes           & Backbone + links & 17 & Distribution conduits \\
Pump            & PU1              & 1  & R2 $\rightarrow$ pressure boost \\
PRV valve       & PRV-1            & 1  & Pressure regulation \\
TCV valve       & TCV-1            & 1  & Flow control \\
{Total pipe length} & {--} & {--} & {16.8 km} \\
{Pipe diameters} & {--} & {--} & {250--300 mm} \\
{Base demand} & {--} & {--} & {59.5 L/s total} \\
\hline
\end{tabular}%
}
\end{table}
The presence of loops and redundant flow paths enables robust simulation of contaminant propagation under varying demand patterns, pump scheduling, and valve configurations. Junction nodes J1--J10 serve both as demand points and as virtual water quality monitoring stations.

\textit{Figure~\ref{fig:network} illustrates the simulated hydraulic network topology.}

\subsection{Reaction Kinetics and Multi-Species Interaction Model}

The EPANET-MSX model governs the temporal evolution of each simulated species through a system of coupled reaction equations{ summarised in Table~\ref{tab:kinetics} and written out below.}
\begin{table}[h!]
\centering
\caption{Multi-species reaction kinetics summary}
\label{tab:kinetics}
\resizebox{\columnwidth}{!}{%
\begin{tabular}{llll}
\hline
\textbf{Species} & \textbf{Sym.} & \textbf{Phase} & \textbf{Kinetics} \\
\hline
Chlorine             & Cl   & Bulk & First-order bulk decay; DO coupling; ClD diagnostic \\
pH                   & pH   & Bulk & First-order relaxation to $\mathrm{pH}_{\mathrm{nat}}$ ($=7.4$) \\
Turbidity            & TURB & Bulk & First-order settling \\
Dissolved Oxygen     & DO   & Bulk & First-order decay; coupled to Cl \\
Conductivity         & COND & Bulk & First-order decay (lumped loss term) \\
Trihalomethanes      & THM  & Bulk & Rate $\propto$ Cl \\
Wall Chlorine Demand & ClD  & Wall & Wall demand ($k_{\mathrm{wall}}\,\mathrm{Cl}$), no feedback \\
\hline
\end{tabular}%
}
\end{table}
Chlorine decay is the central driver of the multi-species system: it simultaneously depletes through intrinsic bulk degradation, oxidative consumption of dissolved oxygen, and biofilm demand on pipe walls. THM formation is modeled as a secondary reaction product whose rate increases with chlorine concentration and residence time, consistent with established DBP formation models in the literature~\cite{ref_thm}.

Written explicitly (all parameter values are listed in the released \texttt{.msx} file and in Table~\ref{tab:msxparams}):
\begin{align}
\frac{d\,\mathrm{Cl}}{dt}   &= -k_{\mathrm{cl}}\,\mathrm{Cl} \;-\; k_{\mathrm{cl,DO}}\,\mathrm{Cl}\,\mathrm{DO}, \label{eq:cl}\\
\frac{d\,\mathrm{ClD}}{dt}  &= k_{w}\,\mathrm{Cl} \quad \text{(wall species, pipes only)}, \\
\frac{d\,\mathrm{THM}}{dt}  &= k_{\mathrm{thm}}\,\mathrm{Cl}, \\
\frac{d\,\mathrm{DO}}{dt}   &= -k_{\mathrm{do}}\,\mathrm{DO} \;-\; k_{\mathrm{cl,DO}}\,\mathrm{Cl}\,\mathrm{DO}, \\
\frac{d\,\mathrm{pH}}{dt}   &= -k_{\mathrm{ph}}\,(\mathrm{pH}-\mathrm{pH}_{\mathrm{nat}}), \\
\frac{d\,\mathrm{TURB}}{dt} &= -k_{\mathrm{turb}}\,\mathrm{TURB}, \\
\frac{d\,\mathrm{COND}}{dt} &= -k_{\mathrm{cond}}\,\mathrm{COND},
\end{align}
where $k_{\mathrm{cl}}$, $k_{\mathrm{do}}$, $k_{\mathrm{turb}}$, $k_{\mathrm{cond}}$, $k_{\mathrm{ph}}$ and $k_{\mathrm{thm}}$ are first-order rate constants (h$^{-1}$), $k_{\mathrm{cl,DO}}$ is the coupling coefficient between chlorine and dissolved oxygen (L\,mg$^{-1}$\,h$^{-1}$), and $k_w$ is the rate at which the biofilm exerts a chlorine demand on the pipe wall (m\,h$^{-1}$). 
ClD is a diagnostic wall-demand species that does not feed back on bulk chlorine. 
The pH is not an algebraic equilibrium species: it is a dynamic species that relaxes at rate $k_{\mathrm{ph}}$ toward the natural pH, $\mathrm{pH}_{\mathrm{nat}}=7.4$. THM is formed in proportion to the residual chlorine, and dissolved oxygen is consumed both by its own decay and by the chlorine-coupled term. Turbidity is removed by first-order settling only (no resuspension), and conductivity decays at first order rather than being treated as a conservative tracer. The same equations, without the wall species $\mathrm{ClD}$, are applied in storage tanks. The first-order bulk term follows~\cite{Rossman1994Chlorine}.

\subsection{Simulation Parameters and Output Dataset}

As summarized in Table~\ref{tab:sim_params}, the EPANET-MSX simulation covers 24 hours with a 1-hour hydraulic step, a 5-minute water quality step and hourly reporting. The demand follows pattern P1, and the water quality state is initialized from reservoirs R1 and R2, with default initial values assigned to all other nodes. Contamination is represented by two pulse injections (chlorine at J3 and turbidity at J7, each lasting two hours) and by localized initial perturbations at J2, J5 and J9. The simulation is exported as 24 hourly states ($t=1$ to $24$\,h) of six species at 13 nodes (78 variables).

\begin{table*}[!t]
\centering
\caption{EPANET-MSX simulation configuration}
\label{tab:sim_params}
\resizebox{\textwidth}{!}{%
\begin{tabular}{ll}
\hline
\textbf{Parameter}              & \textbf{Value} \\
\hline
Simulation duration             & 24 hours \\
Hydraulic time step             & 1\,h \\
Water quality time step         & 5 minutes (MSX \texttt{TIMESTEP} = 300\,s) \\
Reporting interval              & 1\,h \\
Demand pattern                  & P1: 12 hourly multipliers (12-hour cycle, repeated twice over 24\,h) \\
Spin-up discarded               & None (no warm-up period; full 24\,h retained) \\
Source quality (R1, R2)         & Cl 1.0/0.8\,mg/L, pH 7.4, DO 9.0/8.0\,mg/L, COND 380\,$\mu$S/cm, TURB 0.5\,NTU, THM 0 \\
Default initial quality (other nodes) & Cl 0.5\,mg/L, pH 7.4, DO 8.0\,mg/L, COND 380\,$\mu$S/cm, TURB 0.5\,NTU, THM 0 \\
Contamination scenarios         & Pulse injections (\texttt{SETPOINT}): J3 chlorine 5.0\,mg/L, active 06:00--08:00 (2\,h); J7 turbidity 10\,NTU, active 14:00--16:00 (2\,h) \\
Output format                   & One wide CSV (24 rows $\times$ 78 variables, plus a time column) \\
Total synthetic records         & 24 hourly states $\times$ 78 variables (1\,872 values) \\
\hline
\end{tabular}%
}
\end{table*}
\begin{addblock}
\begin{table}[H]
\centering
\caption{Reaction parameters (from the \texttt{.msx} file)}
\label{tab:msxparams}
\resizebox{\columnwidth}{!}{%
\begin{tabular}{llcl}
\hline
\textbf{Symbol} & \textbf{Meaning} & \textbf{Value} & \textbf{Source} \\
\hline
$k_{\mathrm{cl}}$    & Bulk chlorine decay (first order) & 0.10\,h$^{-1}$ & \cite{Rossman1994Chlorine} \\
$k_{\mathrm{cl,DO}}$ & Chlorine--DO coupling coefficient & 0.005\,L\,mg$^{-1}$\,h$^{-1}$ & \cite{Shang2008MSX} \\
$k_{w}$              & Wall chlorine demand rate (biofilm, \texttt{k\_wall}) & 0.02\,m\,h$^{-1}$ & \cite{Rossman1994Chlorine} \\
$k_{\mathrm{thm}}$   & THM formation rate (proportional to Cl) & 0.008\,h$^{-1}$ & \cite{Shang2008MSX} \\
$k_{\mathrm{do}}$    & Dissolved oxygen decay & 0.02\,h$^{-1}$ & \cite{Shang2008MSX} \\
$k_{\mathrm{ph}}$    & pH relaxation rate & 0.01\,h$^{-1}$ & \cite{Shang2008MSX} \\
$\mathrm{pH}_{\mathrm{nat}}$ & Natural (equilibrium) pH & 7.4 & \cite{Shang2008MSX} \\
$k_{\mathrm{turb}}$  & Turbidity settling rate & 0.05\,h$^{-1}$ & \cite{Shang2008MSX} \\
$k_{\mathrm{cond}}$  & Conductivity decay rate & 0.03\,h$^{-1}$ & \cite{Shang2008MSX} \\
\hline
\end{tabular}%
}
\end{table}
\end{addblock}

The resulting synthetic dataset provides complete multi-species time series at all 13 nodes, enabling both spatial analysis of contaminant propagation and temporal anomaly detection. This dataset was subsequently {co-registered} with real IoT measurements to construct the final cyber-physical benchmark used for machine learning model training and evaluation.

\begin{figure*}[h!]
    \centering
    \includegraphics[width=1.85\columnwidth]{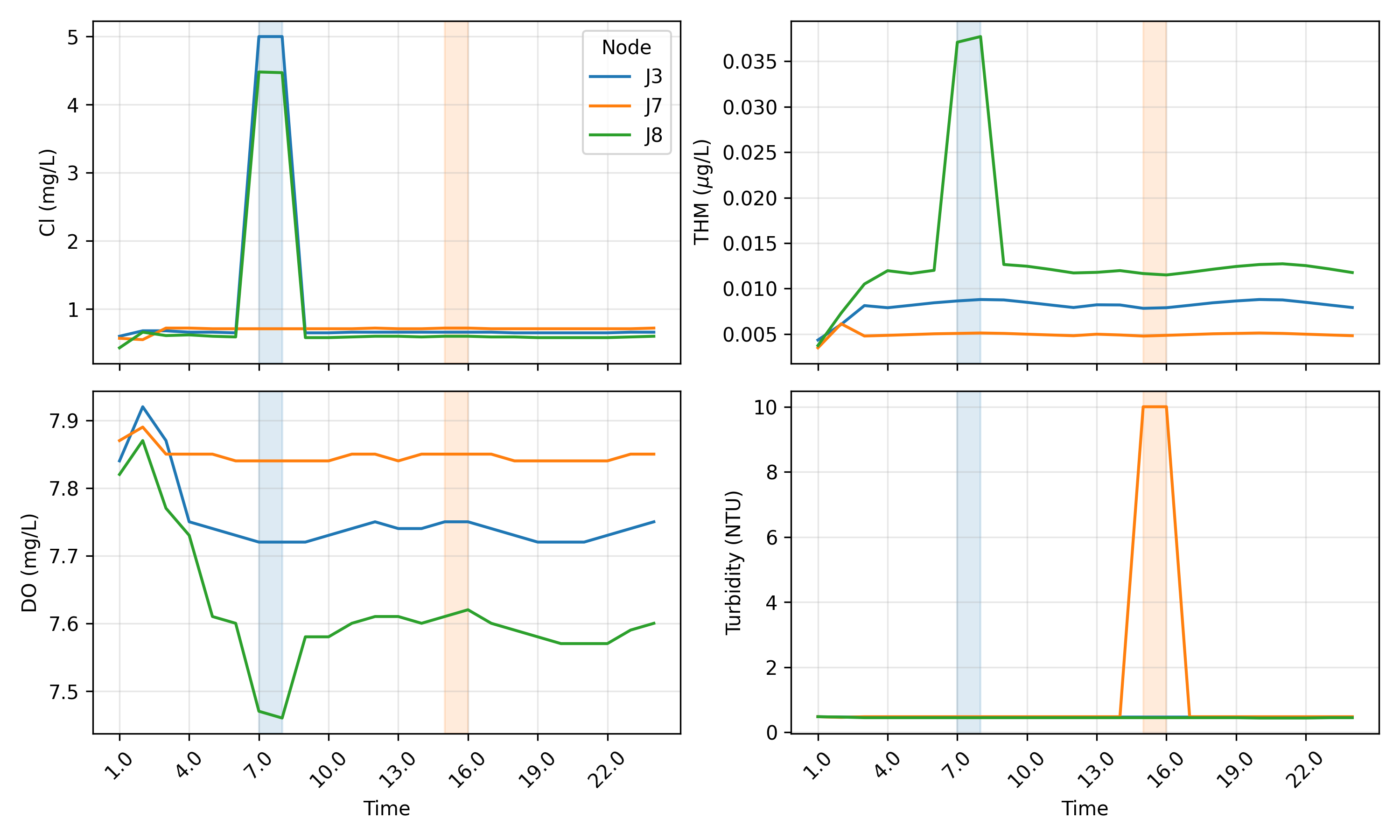}
    \caption{{Simulated concentration profiles of Cl, THM, dissolved oxygen and turbidity at nodes J3 and J7, and at the downstream node J8, following the contamination injection events. Shaded bands mark the injection windows as observed in the hourly report: chlorine at J3 (07:00--08:00, 5.0\,mg/L) and turbidity at J7 (15:00--16:00, 10\,NTU). Background values remain physical (pH\,=\,7.4). The THM response is small (order $10^{-2}$\,$\mu$g/L) and appears mainly downstream of J3, at J8.}}
    \label{fig:msx_timeseries}
\end{figure*}

\textit{Figure~\ref{fig:msx_timeseries} illustrates the simulated concentration profiles of Cl, THM, DO, and Turbidity at nodes J3 and J7 following contamination injection events.}

\section{Data Augmentation and Cyber-Physical Perturbations}\label{sec:perturb}

This section describes the strategies implemented to improve the robustness of the IoT dataset against two categories of challenges: the scarcity of training data and the presence of realistic cyber-physical perturbations. Four complementary axes are explored: statistical and temporal augmentation, synthetic generation through Bootstrapping and TimeGAN, cyber-physical attack modeling, and realistic sensor noise injection.
{All four are implemented in one seeded generator \newline (\texttt{aquasense.perturb}), so any user can regenerate or extend the perturbed data. Augmented and synthetic data are released in separate files and are never mixed into validation or test splits.}

\subsection{Time-Series Augmentation}

Time-series augmentation consists of generating new synthetic observations from real data without altering their essential statistical structure\add{~\cite{Um2017Augment,Iwana2021Augment}}. Five techniques were applied to the columns \texttt{tds\_value}, \texttt{temperature}, \texttt{humidity}, and \newline \texttt{gas\_resistance}.

\textbf{Jittering adds Gaussian noise ($\sigma=0.05$ in the raw units of each channel). Because this amplitude is negligible compared with the gas-sensor resistance (about 90\,000\,$\Omega$), the perturbation is very weak for the gas channel; in the file, it modifies only 19\,\% of the gas rows.}) to each value of the series, mimicking the natural micro-fluctuations of an IoT sensor. \textbf{Scaling} applies a random multiplicative factor within the interval [0.9,1.1] to simulate sensor gain variation. \textbf{Time warping} deforms the time axis through interpolation over random nodes (\texttt{n\_knots = 4}), reproducing real sampling irregularities. \textbf{Window slicing} extracts a 90\% sub-window of the series at a random position, then fills missing values with \texttt{NaN}, simulating partial collection interruptions. Finally, \textbf{Permutation} divides the series into 4 segments and randomly shuffles them, disturbing the temporal order while preserving the overall value distribution.

\begin{table*}[t]
\centering
\footnotesize
\caption{Time-series augmentation techniques}
\label{tab:augmentation}
\resizebox{\textwidth}{!}{%
\begin{tabular}{llll}
\toprule
\textbf{Technique} & \textbf{Parameters} & \textbf{Simulated Effect} & \textbf{{Output file}} \\
\midrule
Jittering & $\sigma = 0.05$ (raw units of each channel) & Sensor micro-fluctuations & {\texttt{augmented\_data.csv}} \\
Scaling & factor $\in [0.9,1.1]$ & Gain variation & {\texttt{augmented\_data.csv}} \\
Time warping & n\_knots = 4 & Sampling irregularities & {\texttt{augmented\_data.csv}} \\
Window slicing & ratio = 0.9 & Random subsequences + NaN & {\texttt{augmented\_data.csv}} \\
Permutation & n\_segments = 4 & Temporal order shuffling & {\texttt{augmented\_data.csv}} \\
\bottomrule
\end{tabular}}
\end{table*}

{\textbf{Why augmented series are not stored as extra columns:} In the first version the five variants were appended as 20 columns aligned row-by-row with the original signals. This creates two problems: (i) a jittered or scaled copy of a feature is almost identical to the feature, so any label computed from the feature can be read off the copy; and (ii) permuted, warped and sliced copies are \emph{misaligned in time} with the row they sit on, so they describe a different moment than the label. Augmentation is therefore now released as separate sequence files that users apply to training data only.}

\subsection{Synthetic Generation: Bootstrapping and TimeGAN}

In addition to direct augmentation, two synthetic generation methods were used to further enrich the volume and diversity of the dataset.

\subsubsection{Bootstrapping}

{Bootstrapping is a controlled statistical generation method consisting
of resampling existing data with replacement to create new observations.
Applied to the IoT time series, we use a standard i.i.d.\ bootstrap in
which individual samples are drawn uniformly with replacement until the
dataset size is doubled (resampling factor of 2). This procedure does not
preserve the temporal order or the short-range autocorrelation of the
original series. It allows reuse of existing data while adding small
variations to the \texttt{tds\_value} feature through Gaussian noise
($\mu = 0$, $\sigma = 0.2$, in raw units) and simulating a slow upward
trend through a linear drift increasing from 0 to 1 over the length of
the synthetic series. The synthetic set is then concatenated with the
original data. This approach represents the baseline level of synthetic
generation: simple to implement, fast to execute, and producing directly usable results.}

\begin{figure}[htbp]
\centering
\includegraphics[width=\columnwidth]{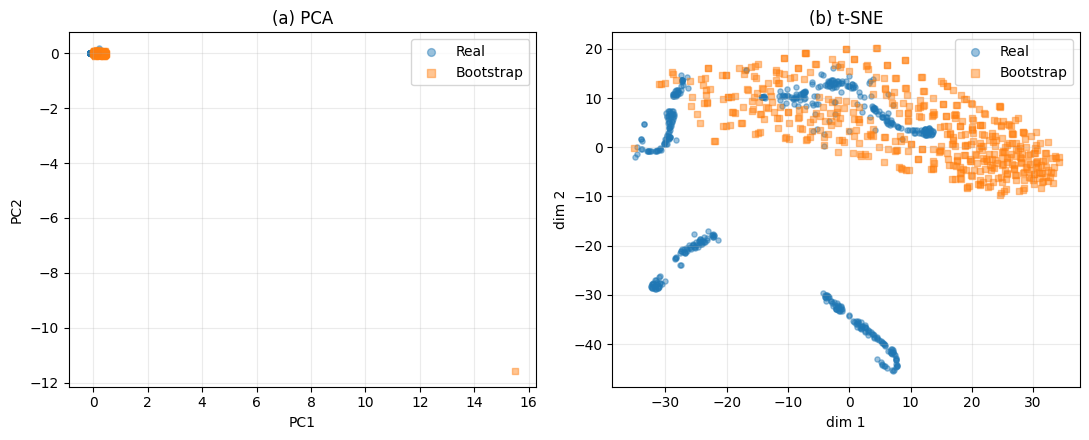}
\caption{PCA (a) and t-SNE (b) overlay of real held-out sequences,
bootstrap sequences and TimeGAN sequences (500 windows of length 24
each).}
\label{fig:pca_tsne}
\end{figure}

\noindent\textbf{Fidelity evaluation:} The qualitative comparison above is replaced by the three standard TimeGAN metrics~\cite{Yoon2019TimeGAN}, computed on held-out real data (Table~\ref{tab:synthfidelity}): the \emph{discriminative score} ($|0.5-\text{accuracy}|$ of a post-hoc classifier separating real from synthetic; lower is better), the \emph{predictive score} (MAE of a one-step forecaster trained on synthetic and tested on real, TSTR; lower is better), and a PCA/t-SNE overlay (Fig.~\ref{fig:pca_tsne}). The simple bootstrap obtains a lower discriminative score than TimeGAN (0.037 vs.\ 0.172).

\begin{table}[H]
\centering
\caption{Fidelity of synthetic sequences (mean $\pm$ std over 5 runs).}
\label{tab:synthfidelity}
\resizebox{\columnwidth}{!}{%
\begin{tabular}{lcc}
\toprule
\textbf{Generator} & \textbf{Discriminative score} $\downarrow$ & \textbf{Predictive score (MAE)} $\downarrow$ \\
\midrule
Real data (reference) & -- & $0.028 \pm 0.000$ \\
Bootstrap & $0.037 \pm 0.031$ & $0.028 \pm 0.000$ \\
TimeGAN & $0.172 \pm 0.047$ & $0.029 \pm 0.000$ \\
\bottomrule
\end{tabular}}
\end{table}

\subsection{Cyber-Physical Attack Modeling}

Five attack types (four classes, including a stealthy variant of false data injection) are simulated, representing the most frequent attack vectors in industrial IoT environments. These attacks are directly injected into time series to create labeled samples for training detection models.

\textbf{False Data Injection (FDI)} modifies transmitted values without affecting the underlying physical system, making falsification invisible to conventional threshold-based alarms\add{~\cite{Liu2011FDI}}. The \textbf{Replay attack} reinjects a past data window to hide an abnormal current state, giving the monitoring system the illusion of normal operation. \textbf{Denial of Service (DoS)} simulates a silent sensor by {interrupting delivery of measurements}. \textbf{Progressive Drift} introduces a growing error, making it particularly difficult to detect without long historical reference, since no fixed threshold can intercept it.

\noindent\textbf{Formal definition.} Let $x_t\in\mathbb{R}$ be a clean channel, $\sigma_x$ its standard deviation on the training split, and $\mathcal{A}=[t_s,t_s+L)$ an attack interval. The attacked signal $\tilde{x}_t$ equals $x_t$ outside $\mathcal{A}$ and, for $t\in\mathcal{A}$:
\begin{align}
\text{FDI (bias):}\quad & \tilde{x}_t = x_t + \beta\,\sigma_x,\quad \beta\sim\mathcal{U}\{\pm[0.5,2]\},\\
\text{FDI (stealthy):}\quad & \tilde{x}_t = \mathrm{clip}\big(x_t + \beta\,\sigma_x,\ \ell,\ u\big),\\
\text{Replay:}\quad & \tilde{x}_t = x_{t-\Delta},\quad \Delta\ge L,\ [t_s{-}\Delta,t_s{-}\Delta{+}L)\ \text{attack-free},\\
\text{DoS:}\quad & \tilde{x}_t = x_{t_s-1}\ \ \text{(frozen)}\quad\text{or}\quad \tilde{x}_t=\texttt{NaN}\ \ \text{(dropped)},\\
\text{Drift:}\quad & \tilde{x}_t = x_t + \gamma\,\sigma_x\,\frac{t-t_s}{L},\quad \gamma\sim\mathcal{U}[1,3],
\end{align}
where $[\ell,u]$ is the normal operating range of the attacked channel. Onsets $t_s$ are drawn uniformly and intervals never overlap. The released stream contains 14 attack segments (replay 7, drift 3, DoS 2, FDI 1, stealthy FDI 1) of 52 to 287 samples each, and 10.0\% of the real samples are attacked. Each attack type acts on specific channels (Table~\ref{tab:attackchannels}): FDI on \texttt{Cl\_J7}, stealthy FDI on \texttt{TURB\_J7} and DoS on \texttt{TURB\_J3}, so these three types never alter a real sensor channel, whereas replay and drift act on real and simulated channels. Both DoS variants are implemented: one interval is frozen (last valid value repeated) and one is dropped (\texttt{NaN}, 271 samples on \texttt{TURB\_J3}). Each sample carries \texttt{attack\_type} $\in\{$none, FDI, FDI-stealth, replay, DoS, drift$\}$, \texttt{attack\_id} and \texttt{attack\_channel}. The earlier zero-forcing DoS was dropped because a real DoS produces missing or stale packets, not zeros, and zeros are trivially detectable.

\begin{table}[H]
\centering
\caption{Attack and fault types in the real stream: number of segments, length and channels affected (number of samples).}
\label{tab:attackchannels}
\footnotesize
\begin{tabular}{llp{4.6cm}}
\toprule
\textbf{Type} & \textbf{Segments (length)} & \textbf{Channels (samples)} \\
\midrule
FDI & 1 (102) & Cl\_J7 (102) \\
FDI-stealth & 1 (265) & TURB\_J7 (265) \\
DoS & 2 (244--271) & TURB\_J3 (515) \\
Replay & 7 (52--287) & humidity (287), TURB\_J7 (201), TURB\_J3 (169), gas (159), TDS (143), temp. (85) \\
Drift & 3 (54--196) & TDS (388), humidity (54) \\
\midrule
White noise & 2{,}368 samples & all channels simultaneously \\
Spike & 118 samples & temp. (22), gas (18), humidity (15), Cl\_J7 (14), TURB\_J3 (14), TDS (12), TURB\_J7 (12), Cl\_J3 (11) \\
Extreme & 47 samples & temp. (8), TURB\_J3 (8), Cl\_J3 (7), Cl\_J7 (7), TURB\_J7 (5), gas (5), TDS (5), humidity (2) \\
\bottomrule
\end{tabular}
\end{table}

The attacks remain close to the normal range of the observed channels. FDI samples on \texttt{Cl\_J7} lie between 0.393 and 0.480\,mg/L (mean 0.437), against a clean maximum of 0.476; stealthy FDI and replay samples on \texttt{TURB\_J7} lie between 0.417 and 0.737\,NTU, inside the clean range (0.350--0.821); DoS and replay samples on \texttt{TURB\_J3} lie between 0.252 and 0.549\,NTU (clean range 0.203--0.606). The frozen DoS values never produce a zero reading.

\begin{table*}[t]
\centering
\footnotesize
\caption{Cyber-physical attack scenarios}
\label{tab:attacks}
\resizebox{\textwidth}{!}{%
\begin{tabular}{llll}
\toprule
\textbf{Attack Type} & \textbf{{Parameters}} & \textbf{Observable Effect} & \textbf{Detection Difficulty} \\
\midrule
False Data Injection & {$\beta\in\pm[0.5,2]\,\sigma_x$} & {Step offset} & High \\
{FDI (stealthy)} & {as FDI, clipped to $[\ell,u]$} & {In-range offset} & {Very high} \\
Replay &{$\Delta\ge L$} & {Plausible but stale values} & Moderate \\
Denial of Service &{frozen or dropped} & {Flat line or missing samples} & Low \\
Drift & {$\gamma\in[1,3]\,\sigma_x$ over $L$} & {Slowly growing bias} & Very high \\
\bottomrule
\end{tabular}}
\end{table*}

\subsection{Realistic Noise and Anomalies}

Three types of perturbations are injected to simulate faulty behaviors of physical sensors under real operating conditions. These perturbations are distinct from intentional attacks: they reproduce unintentional failures linked to environmental conditions, hardware wear, or electronic interference{, and they carry a separate label (\texttt{fault\_type}) so that models can be tested on the practically important question of distinguishing faults from attacks}.

\textbf{White noise }(\(\mathcal{N}(0,0.2^2)\), in standardized z-score units) models small, zero-mean Gaussian fluctuations commonly observed in IoT sensor measurements. This perturbation is added to legitimate measurements to evaluate the model's robustness to random measurement variability without incorrectly identifying minor fluctuations as significant anomalies. 

\textbf{Spikes} are generated by adding isolated high-amplitude impulses (\(+5\sigma_x\)) at randomly selected time indices, simulating transient sensor glitches or erroneous readings potentially caused by electromagnetic interference or acquisition errors. 

\textbf{Extreme values} are generated by replacing selected measurements with the corresponding sensor's full-scale limit, simulating saturation or temporary sensor malfunctions. Gaussian white noise is independently injected into all sensor channels, whereas spikes and extreme-value perturbations are introduced into one channel at a time to assess the model's sensitivity to localized anomalies.

\begin{table*}[htbp]
\centering
\scriptsize
\caption{Realistic sensor noise and anomaly scenarios}
\label{tab:noise}
\resizebox{\textwidth}{!}{%
\begin{tabular}{llll}
\toprule
\textbf{Noise Type} & \textbf{{Parameters}} & \textbf{Frequency} & \textbf{Detection Difficulty} \\
\midrule
White noise & $\mathcal{N}(0,0.2\sigma_x)$, 10.0\% of samples (2{,}368) & Very frequent & Low \\
Spike / Impulse & $+5\sigma_x$, 0.5\% of samples (118) & Occasional & Moderate \\
Extreme values & full-scale value, 0.2\% of samples (47) & Rare but critical & High \\
\bottomrule
\end{tabular}}
\end{table*}

These three combined types cover common real IoT sensor failures, from normal micro-variations to complete faults.

\subsection{Section Summary}

All strategies deployed in this section form a coherent and complete dataset robustness pipeline structured over four increasing levels of complexity.

\begin{table*}[t]
\centering
\caption{Comprehensive summary of data augmentation and cyber-physical perturbation strategies}
\label{tab:synthesis}
\begin{tabularx}{\textwidth}{L{0.75} L{1.10} L{1.30} L{0.85}}
\toprule
\textbf{Axis} & \textbf{Methods used} & \textbf{Output} & \textbf{Objective} \\
\midrule
Temporal augmentation & Jittering, scaling, time warping, window slicing, permutation
  & {5 separate training-only files}
  & Diversify the training dataset \\
\addlinespace
Synthetic generation & Bootstrapping, TimeGAN
  & Realistic synthetic series {(separate files, fidelity in Table~\ref{tab:synthfidelity})}
  & Cover rare scenarios and increase volume \\
\addlinespace
Cyber-physical attacks & FDI, {stealthy FDI,} replay, DoS, drift
  & {5} labeled attack {types, sample-level labels}
  & Train intrusion detection models \\
\addlinespace
Realistic sensor noise & White noise, spikes, extreme values
  & 3 perturbation types {(\texttt{fault\_type})}
  & Simulate unintentional sensor failures \\
\bottomrule
\end{tabularx}
\end{table*}

\section{Temporal Fusion and Water Quality Labeling}\label{sec:fusion}

This section presents the final construction of the supervised dataset used for training classification models. It includes temporal alignment of heterogeneous data sources, a fusion strategy, and an automatic water quality labeling procedure based on standardized physicochemical indicators.

\subsection{Temporal Alignment}

\subsubsection{Context and Objective}

The two data sources, real physical sensors {and} hydraulic simulation outputs, use heterogeneous time formats. To enable reliable fusion, a common temporal key expressed in the same unit is required.

\noindent\textbf{Alignment.} Both sources are indexed by elapsed seconds since the start of their record, $\tau$. The real record runs from 2025-05-19 12:39:08 to 2025-05-20 14:01:00, i.e.\ about 25.4\,h, and contains 23{,}688 samples (median step 1\,s, mean step 3.9\,s); it is not continuous, since 23 gaps longer than 60\,s interrupt it, the longest lasting about 10.1\,h. The simulation file covers 24\,h (\texttt{time\_h} from 1.00 to 24.00\,h, one state per hour), so no spin-up period is discarded. The start of the real record is aligned with the start of the simulation clock:
\begin{equation}
\tau_{\mathrm{sim}} = \tau_{\mathrm{real}} + \Delta_0,\qquad \Delta_0 = 0\ \mathrm{s}.
\end{equation}

\subsubsection{Processing of Real Sensor Data}

The \texttt{timestamp} field in the real dataset follows the ISO 8601 datetime format (e.g., \texttt{2025-05-19 12:39:08}). It is converted as follows:

\begin{equation}
{\tau_{\mathrm{real}} = t - t_0\ \ [\mathrm{s}],\quad t_0=\text{first timestamp}}
\end{equation}

Rows containing invalid or missing timestamps are removed using \texttt{dropna()} before conversion, ensuring the integrity of the temporal join key.

\subsubsection{Processing of Simulated Data}

The EPANET-MSX export (\texttt{simulated\_data.csv}) is already in wide format: one row per hourly reporting step (24 rows) and one column per species and node (6 species $\times$ 13 nodes = 78 columns, e.g.\ \texttt{Cl\_J3}). Time is stored as a numeric column \texttt{time\_h} (elapsed hours, from 1.00 to 24.00\,h). The preprocessing steps are:

\begin{itemize}
    \item Removing rows with a missing \texttt{time\_h};
    \item Converting hours to elapsed seconds:
\end{itemize}

\begin{equation}
\tau_{\mathrm{sim}} = \mathrm{round}\!\left(3600\,\texttt{time\_h}\right)\ \ [\mathrm{s}]
\end{equation}

\begin{itemize}
    \item Applying the clock offset $\Delta_0 = 0$ defined above; no spin-up rows are discarded since the file covers only 24\,h.
\end{itemize}

\noindent If a simulation file is delivered in long format (\texttt{Time} as \texttt{HH:MM}, one row per node), the pipeline also supports it: $\tau_{\mathrm{sim}} = 3600\,\mathrm{Hour} + 60\,\mathrm{Minute}$, followed by a pivot to wide format.

\subsubsection{Time Format Correspondence}

\begin{table}[H]
\centering
\caption{Temporal format correspondence across data sources}
\label{tab:time_formats}
\renewcommand{\arraystretch}{1.2}
\resizebox{\columnwidth}{!}{%
\begin{tabular}{lll}
\toprule
\textbf{Source} & \textbf{Original Format} & \textbf{Temporal Key} \\
\midrule
\texttt{Real\_data.csv} & ISO 8601 timestamp & elapsed s, $\tau_{\mathrm{real}}$ \\
\texttt{simulated\_data.csv} & \texttt{time\_h} (numeric hours) & $\mathrm{round}(3600\,\texttt{time\_h})$ \\
Merged dataset & Integer (\texttt{tau\_s}) & Common join key \\
\bottomrule
\end{tabular}}
\end{table}

\subsection{Data Fusion Strategy}

\subsubsection{Join Selection}

Each real sample is matched with at most one simulated network state through an as-of join with a tolerance of one reporting interval (3{,}600\,s):
\begin{verbatim}
df = pd.merge_asof(real.sort_values("tau_s"),
             sim_wide.sort_values("tau_s"),
             on="tau_s", direction="backward",
             tolerance=3600)
\end{verbatim}
Simulated columns whose name already exists in the fused table (\texttt{Cl\_J3}, \texttt{TURB\_J3}, \texttt{Cl\_J7}, \texttt{TURB\_J7}) are renamed with the suffix \texttt{\_epanet}. The column \texttt{sim\_state\_h} records the hour of the simulated state matched to each real sample. The fused table has exactly one row per real sample (23{,}688 rows), with no duplication.

\noindent\textbf{Observed Cl and TURB channels.} The columns \texttt{Cl\_J3}, \newline \texttt{TURB\_J3}, \texttt{Cl\_J7} and \texttt{TURB\_J7} are the channels on which the attack and fault generator acts. They are noisy signals around constant baselines and are distinct from the hourly EPANET outputs (\texttt{*\_epanet}). On the 18{,}787 real samples without attack or fault, their mean (standard deviation) is 0.500 (0.042) for \texttt{Cl\_J3} and 0.349 (0.042) for \texttt{Cl\_J7} (mg/L), and 0.401 (0.067) for \texttt{TURB\_J3} and 0.599 (0.083) for \texttt{TURB\_J7} (NTU). Over all real samples, fault injection widens these ranges to 0--5\,mg/L and 0--1000\,NTU: 27 \texttt{extreme} samples set one of these channels to zero (11 samples) or to its full-scale value (16 samples).

\subsubsection{Temporal Propagation}

Simulated data are produced every hour, whereas real sensors provide measurements every 1--2\,s (median step 1\,s). The backward as-of join assigns to each real sample the most recent simulated state within 3{,}600\,s, the EPANET-MSX reporting resolution. Applied to the 23{,}688 real samples, the join gives a simulated state to 20{,}795 samples (87.8\%). The remaining 2{,}893 samples keep missing values (NaN) for the simulated columns: 1{,}800 samples recorded during the first hour, before the first simulated state at $\tau=1$\,h, and 1{,}093 samples recorded more than one hour after the last state (the real record lasts 25.4\,h, the simulation 24\,h). Because the real record is interrupted by long gaps (Section~\ref{sec:acq}), only 11 of the 24 hourly states are matched by at least one real sample, and a single state is shared by at most 2{,}816 consecutive samples. As a consequence, the chlorine pulse at J3 (state at $\tau=7$\,h, matched by 1{,}427 samples) is present in the fused table, whereas the turbidity pulse at J7 (states at $\tau=15$ and 16\,h) is not matched by any real sample and remains visible only in \texttt{simulated\_data.csv}. Elapsed time $\tau$ is counted from the start of each record, so $\tau=7$\,h corresponds to about 19:39 on 19 May in the real record, not to 07:00 clock time. No value is invented and no value crosses a node boundary. The approach assumes quasi-stationary behavior of the network over one hour.

\subsubsection{Synthetic Sequences}

The TimeGAN sequences have no common clock and no associated simulated state. The 23{,}688 synthetic rows are appended to the fused table with \texttt{origin = timegan}, \texttt{attack\_type = fault\_type = none}, and missing values (NaN) for columns not provided by TimeGAN. Only \texttt{tds\_value}, \texttt{temperature}, \texttt{humidity}, and \texttt{gas\_resistance} are shared with the real data. These synthetic sequences are used for training only. The resulting file contains 47{,}376 rows and 97 columns, of which 23{,}688 have \texttt{origin = real}. The TimeGAN rows carry the four sensor channels only; their labels \texttt{quality\_final} and \texttt{wq\_class} are missing, and \texttt{attack\_type} and \texttt{fault\_type} are \texttt{none}.

\subsubsection{Post-Fusion Cleaning}

\begin{itemize}
    \item Removed columns: \texttt{Time}, \texttt{Hour}, \texttt{Minute}, \texttt{time\_h}, \texttt{tau\_s}, \texttt{tau\_real}, \texttt{quality\_score} and the twenty augmented copies of the sensor channels \newline (released separately in \texttt{augmented\_data.csv});
    \item Preserved column: \texttt{timestamp};
    \item Metadata columns \texttt{origin}, \texttt{attack\_id}, \texttt{attack\_channel}, \texttt{fault\_channel}, \texttt{sim\_state\_h};
    \item Labels: \texttt{attack\_type}, \texttt{fault\_type}, \texttt{wq\_score}, \texttt{wq\_class} and, as last column, the legacy label \texttt{quality\_final}.
\end{itemize}

In the final file, \texttt{attack\_type} contains five attack classes (replay: 1{,}044; DoS: 515; drift: 442; FDI-stealth: 265; FDI: 102), representing 10.0\,\% of the 23{,}688 real samples. The column \texttt{fault\_type} contains three fault classes (white noise: 2{,}368; spike: 118; extreme: 47), representing 10.7\,\% of the real samples. The dataset comprises 23{,}688 real rows and 23{,}688 TimeGAN-generated rows.

\subsection{Automatic Water Quality Label Generation}\label{sec:labels}

\noindent\textbf{Label families.} The released table provides (i) \texttt{wq\_score} and \texttt{wq\_class}, computed only from water parameters with fixed regulatory thresholds: pH, turbidity, residual chlorine, THM and conductivity at node J3 from the simulation, and TDS from the raw sensor record; (ii) \texttt{attack\_type} and \texttt{fault\_type}, from the perturbation generator (Section~\ref{sec:perturb}); and (iii) \texttt{quality\_final}, a legacy label inherited from the first version of the dataset, derived from nine inputs that include ambient variables (air temperature, humidity, gas resistance). Air humidity and VOC level do not describe the water, so \texttt{quality\_final} is kept for backward compatibility only and is not used as a benchmark target.

\subsubsection{Composite Scoring Approach}

Automatic labeling is based on a composite score computed from {six physicochemical} indicators. Each parameter is evaluated using two thresholds:

\begin{itemize}
    \item 2 points: fully compliant
    \item 1 point: acceptable
    \item 0 point: non-compliant
\end{itemize}

The total score ranges from 0 to {12} points (the maximum reached in the dataset is 11).

\subsubsection{Indicators and Thresholds}

The composite quality score is derived from six physicochemical indicators evaluated at a single reference node of the network (J3). For each parameter, two evaluation levels are defined: an optimal range corresponding to full compliance (2 points) and an acceptable range corresponding to partial compliance (1 point), while values outside these limits receive 0 points. Table~\ref{tab:quality_thresholds} summarizes the selected thresholds and their reference standards.

At node J3 the simulated indicators take the values reported in Table~\ref{tab:stats_descriptives}: pH 7.4, turbidity 0.46--0.48\,NTU, THM below 0.01\,$\mu$g/L, conductivity 363--370\,$\mu$S/cm, and chlorine between 0.60 and 5.00\,mg/L. The TDS indicator is computed on the raw record, before fault injection.

On the 23{,}688 real rows, \texttt{tds\_value} has a median of 0.362 and a mean of 0.536, with values from 0.007 to 1.196 and no negative value. These values do not appear to be expressed in ppm (no voltage column is released, so the conversion cannot be verified), and the ppm thresholds of Table~\ref{tab:quality_thresholds} cannot be meaningfully applied to them: since every value is below 300, the TDS indicator always receives the maximum score of 2 points and does not discriminate between samples. The simulated variables follow the units declared in the MSX file: THM in $\mu$g/L, turbidity in NTU, residual chlorine and dissolved oxygen in mg/L, and conductivity in $\mu$S/cm.

\begin{table*}[t]
\centering
\caption{Water quality indicators and scoring thresholds}
\label{tab:quality_thresholds}
\renewcommand{\arraystretch}{1.15}
\begin{tabular}{lllll}
\toprule
\textbf{Parameter} & \textbf{Optimal Range} & \textbf{Acceptable Range} & \textbf{Score} & \textbf{Reference} \\
\midrule
pH & 6.5--8.5 & 6.0--9.0 & 2 / 1 / 0 & \cite{WHO2022,EUDirective2020} \\
Turbidity (NTU) & $\leq 1$ & $\leq 5$ & 2 / 1 / 0 & \cite{WHO2022} \\
Residual Chlorine (mg/L) & 0.2--0.5 & 0.1--5.0 & 2 / 1 / 0 & \cite{WHO2022} \\
Trihalomethanes ($\mu$g/L) & $\leq 60$ & $\leq 100$ & 2 / 1 / 0 & \cite{EUDirective2020} \\
Conductivity ($\mu$S/cm) & $\leq 800$ & $\leq 2500$ & 2 / 1 / 0 & \cite{EUDirective2020} \\
TDS (mg/L) & $\leq 300$ & $\leq 600$ & 2 / 1 / 0 & \cite{WHO2022} \\
\bottomrule
\end{tabular}
\end{table*}

\subsubsection{\texorpdfstring{{Class Assignment with Fixed Thresholds}}{Class Assignment with Fixed Thresholds}}

Classes are assigned with fixed rules, including a health-based override (Table~\ref{tab:quality_classes}). Applied to the 20{,}795 real samples that have a simulated state, with the TDS indicator computed before fault injection, the rules give 20{,}795 ``Very Good'' samples and no ``Good'' or ``Poor'' sample: the score equals 11 for every sample. Three reasons explain this. The simulated J3 values stay within regulatory limits (pH 7.4, turbidity 0.46--0.48\,NTU, THM below 0.009\,$\mu$g/L, conductivity 363--370\,$\mu$S/cm). The TDS values (maximum 1.196) always receive the maximum score. And chlorine (0.60--5.00\,mg/L) never falls in the optimal range of 0.20--0.50\,mg/L, hence a score of 11 rather than 12, while the chlorine pulse reaches exactly 5.0\,mg/L in 1{,}427 samples without exceeding the critical threshold of $>5$\,mg/L. The remaining 2{,}893 real samples have no simulated state and receive no class. We consequently do not use \texttt{wq\_class} as a benchmark target.

\begin{table}[H]
\centering
\caption{Final water quality classes}
\label{tab:quality_classes}
\renewcommand{\arraystretch}{1.2}
\resizebox{\columnwidth}{!}{%
\begin{tabular}{lll}
\toprule
\textbf{Class} & \textbf{Condition} & \textbf{Interpretation} \\
\midrule
Very Good & $S\ge 11$ and no critical violation & High-quality compliant water \\
Good & $8\le S<11$ and no critical violation & Acceptable water quality \\
Poor & $S<8$ \emph{or} any critical violation$^{*}$ & Potential health risk \\
\bottomrule
\multicolumn{3}{l}{$^{*}$Cl $<0.1$ or $>5$\,mg/L, TURB $>5$\,NTU, or THM $>100\,\mu$g/L.}
\end{tabular}}
\end{table}

\subsubsection{Final Dataset Structure}

The final dataset contains:

\begin{itemize}
    \item Input features: the four sensor channels, the four observed Cl/TURB channels and the 78 simulated variables;
    \item Targets: \texttt{attack\_type}, \texttt{fault\_type}, \texttt{wq\_class} (constant, see above) and the legacy \texttt{quality\_final};
    \item 47{,}376 rows: 23{,}688 real and 23{,}688 TimeGAN rows;
    \item Missing values: the simulated variables are missing for the 2{,}893 real samples without a matched state and for all TimeGAN rows.
\end{itemize}

\subsection{Structure of the Dataset}

The release is organised as linked tables rather than one wide file (Table~\ref{tab:schema}); the measured tables are indexed by \texttt{timestamp} and the simulated table by elapsed time \texttt{time\_h}.

\begin{table*}[t]
\centering
\footnotesize
\caption{Structure of the released dataset (v1.0).}
\label{tab:schema}
\renewcommand{\arraystretch}{1.15}
\begin{tabular}{p{3.6cm}rrp{8.4cm}}
\toprule
\textbf{File} & \textbf{Rows} & \textbf{Cols} & \textbf{Content} \\
\midrule
\texttt{Real\_data.csv} & 23{,}688 & 6 & \texttt{timestamp}, temperature ($^\circ$C), humidity (\%), gas resistance, \texttt{tds\_value} and {qualite or quality\_final: indicate the real 6th column}. \\
\texttt{simulated\_data.csv} & 24 & 79 & \texttt{time\_h} (hourly, 24\,h), followed by 78 water-quality variables (Cl, pH, TURB, DO, COND, and THM) across 13 nodes in wide format. \\
\texttt{augmented\_data.csv} & 23{,}688 & 27 & Measured sensor data with \texttt{qualite}, and five augmented variants of each of the four sensor channels. \\
\texttt{synthetic\_data.csv} & 23{,}688 & 4 & Synthetic sensor channels: \texttt{tds\_value}, \texttt{temperature}, \texttt{humidity}, and \texttt{gas\_resistance}. \\
\texttt{fused\_dataset.csv} & 47{,}376 & 97 & Fused real and synthetic data, including sensor and simulated water-quality variables, metadata (\texttt{origin}, \texttt{attack\_id}, \texttt{attack\_channel}, \texttt{fault\_channel}), \texttt{attack\_type}, \texttt{fault\_type}, \texttt{sim\_state\_h}, \texttt{wq\_score}, \texttt{wq\_class}, and \texttt{quality\_final}. The augmented copies are not included. \\
\texttt{network/water\_network.inp} & -- & -- & EPANET hydraulic network model with 10 junctions, 2 reservoirs, 1 tank, 17 pipes, 1 pump, 2 valves, and 4 water-quality sources. The model uses a 24-hour simulation duration, hourly hydraulic and quality timesteps, and L/s flow units. \\
\texttt{network/water\_quality.msx} & -- & -- & EPANET-MSX multi-species model defining six bulk water-quality species (Cl, pH, TURB, DO, COND, and THM), one wall species (ClD), reaction coefficients and terms, initial/global quality conditions, and two configured setpoint sources associated with attack patterns at J3 and J7. \\
\texttt{network/injection\_schedule.csv} & -- & -- & Injection schedule. \\
\texttt{splits.json} & -- & -- & Train/validation/test indices for each task and split protocol. \\
\bottomrule
\end{tabular}
\end{table*}

\subsection{Descriptive Statistics per Variable}

Table~\ref{tab:stats_descriptives} summarizes the descriptive statistics
of the four IoT sensor variables, computed on the 23{,}688 real rows of
\texttt{Real\_data.csv}, and of six water-quality variables at node J3,
computed on the 24 hourly states of \texttt{simulated\_data.csv}. The
TimeGAN rows of the fused dataset are excluded, since they are generated
and not measured.

\begin{table}[h!]
\centering
\caption{Descriptive statistics of selected variables. Sensor statistics
are computed on the 23{,}688 real rows; simulated statistics on the 24
hourly states at node J3.}
\label{tab:stats_descriptives}
\setlength{\tabcolsep}{3pt}
\renewcommand{\arraystretch}{1.1}
\small
\resizebox{\columnwidth}{!}{%
\begin{tabular}{lrrrrr}
\toprule
\textbf{Variable} & \textbf{Min} & \textbf{Max} &
\textbf{Mean} & \textbf{Std.} & \textbf{Missing (\%)} \\
\midrule
\multicolumn{6}{l}{\textit{IoT Sensors (real rows, $n=23{,}688$)}} \\
\midrule
Air temperature ($^\circ$C)$^{a}$ & 25.30 & 37.32 & 30.75 & 2.22 & 0.00 \\
Relative humidity (\%RH)$^{a}$ & 35.07 & 73.60 & 47.98 & 7.30 & 0.00 \\
Gas resistance ($\Omega$)$^{a}$ & 670.00 & 226,884.00 & 95,227.08 & 43,108.05 & 0.00 \\
TDS (unit unverified)$^{b}$ & 0.01 & 1.20 & 0.54 & 0.40 & 0.00 \\
\midrule
\multicolumn{6}{l}{\textit{Water Distribution Network, node J3 (EPANET-MSX, $n=24$)}} \\
\midrule
pH & 7.40 & 7.40 & 7.40 & 0.00 & 0.00 \\
Turbidity (NTU)$^{c}$ & 0.46 & 0.48 & 0.47 & 0.01 & 0.00 \\
Residual chlorine (mg/L)$^{c,d}$ & 0.60 & 5.00 & 1.02 & 1.23 & 0.00 \\
Dissolved oxygen (mg/L)$^{c}$ & 7.72 & 7.92 & 7.75 & 0.05 & 0.00 \\
Trihalomethanes ($\mu$g/L)$^{c}$ & 0.0043 & 0.0088 & 0.0081 & 0.0010 & 0.00 \\
Conductivity ($\mu$S/cm)$^{c}$ & 362.91 & 370.12 & 364.40 & 1.63 & 0.00 \\
\bottomrule
\multicolumn{6}{p{1.05\columnwidth}}{\footnotesize
$^{a}$Temperature is recorded in $^\circ$C, relative humidity in \%RH
and gas resistance in $\Omega$ by the BME680. These are ambient (air)
measurements.

$^{b}$The TDS values are not in ppm (see Section~\ref{sec:calib}); the
unit is left unspecified because the raw probe voltage was not retained.

$^{c}$Units follow the MSX species declarations: turbidity in NTU,
residual chlorine and dissolved oxygen in mg/L, conductivity in
$\mu$S/cm and trihalomethanes in $\mu$g/L.

$^{d}$The maximum of 5.00\,mg/L corresponds to the chlorine pulse
injected at J3 (reported at 07:00 and 08:00). The wall species ClD is
internal to the MSX model and is not reported.
}
\end{tabular}}
\end{table}

On the real rows, the four sensor channels contain no missing value and no physically impossible reading: relative humidity never exceeds 100\% (maximum 73.6\%), and no gas resistance or TDS value is negative. The sensor columns of the fused table are the \emph{observed} channels after fault and attack injection; rows without fault or attack are identical to the raw record. The extreme values found in them are injected faults flagged by \texttt{fault\_type}: eight full-scale temperature values ($-40$ or $85\,^\circ$C) and three TDS values of 1000 (\texttt{extreme}), one humidity above 100\% (\texttt{spike}), and 37 negative gas resistances produced by white noise added to low readings.

Two features of the real stream deserve attention. First, the mean air
temperature (30.75\,$^\circ$C) lies above the normal range of
Table~\ref{tab:thresholds} (10--30\,$^\circ$C), and 62\% of the real
samples fall outside it; the temperature-based ambient flag is therefore
active for most of the record and should be read as a station-specific
offset rather than as an anomaly. Second, gas resistance falls below
8{,}000\,$\Omega$ in 818 samples (3.5\%), grouped in 38 short episodes
concentrated in five sessions of the record, with a minimum of 670\,$\Omega$.

The simulated variables are complete (no missing value) and vary over
the 24\,h horizon. Chlorine shows the largest variation, driven by the
injected pulse, whereas pH remains at its natural value of 7.4, and
turbidity, dissolved oxygen, THM and conductivity vary only slightly at
J3, since the turbidity pulse is applied at J7 and not at J3.

\subsection{Distribution of Target Classes}

Table~\ref{tab:distribution_classes} reports the water-quality class \texttt{wq\_class} of the real rows. The TimeGAN rows carry no simulated state and no class.

\begin{table}[h!]
\centering
\caption{Distribution of \texttt{wq\_class} on the real rows.}
\label{tab:distribution_classes}
\begin{tabular}{lcc}
\toprule
\textbf{Class} & \textbf{Samples} & \textbf{Share (\%)} \\
\midrule
Very Good & 20{,}795 & 87.8 \\
Good      & 0        & 0.0 \\
Poor      & 0        & 0.0 \\
No simulated state & 2{,}893 & 12.2 \\
\midrule
\textbf{Total} & \textbf{23{,}688} & \textbf{100.0} \\
\bottomrule
\end{tabular}
\end{table}

The legacy label \texttt{quality\_final} (real rows: 1{,}578 Very Good, 5{,}154 Good, 16{,}956 Poor) is unrelated to \texttt{wq\_class}: among the 20{,}795 real samples with a simulated state, it is Poor for 14{,}970 (72.0\%), Good for 4{,}441 (21.4\%) and Very Good for 1{,}384 (6.7\%), whereas \texttt{wq\_class} is Very Good for all of them.

\section{Exploratory Data Analysis}\label{sec:eda}

\subsection{Correlation Analysis}\label{subsec:correlation}

The Spearman rank-correlation matrix computed over the main physicochemical parameters reveals six associations ($|\rho|>0.40$), all of them among simulated species. Simulated quantities are summarised by their time-averaged value at each of the 13 network nodes (J1--J10, R1, R2, T1) over the 24 hourly states, since their temporal variation within a node is not negligible (mean standard deviation per node: 0.24\,mg/L for chlorine, 0.52\,NTU for turbidity, 5.5\,$\mu$S/cm for conductivity); 95\% confidence intervals are obtained by node-level bootstrap (5000 resamples). Sensor channels are correlated on the 18{,}787 real rows that carry neither an attack nor a fault, so that injected perturbations do not affect the coefficients. We report effect sizes and intervals rather than $p$-values. Augmented variants are not part of the released table and are excluded. The significant pairs are summarised in Table~\ref{tab:spearman_corr}.

\textbf{Scope and limits.} (i)~\texttt{pH} is constant across nodes (7.4) and is excluded. (ii)~Correlations between real and simulated channels are not computed: they are co-registered, not coupled (Section~\ref{sec:fusion}). (iii)~No sensor pair reaches the threshold; the strongest, \texttt{humidity}~$\leftrightarrow$~\texttt{temperature}, is only $\rho=+0.29$, followed by \texttt{tds\_value}~$\leftrightarrow$~\texttt{temperature} ($-0.13$) and \texttt{gas\_resistance}~$\leftrightarrow$~ \newline \texttt{temperature} ($-0.12$). (iv)~With $n=13$ nodes, the coefficients describe the spatial structure of the simulated network, not independent samples.

\definecolor{verystrongbg}{RGB}{198,224,180}
\definecolor{moderatestrongbg}{RGB}{255,242,204}
\definecolor{moderatebg}{RGB}{222,235,247}
\definecolor{headerblue}{RGB}{31,73,125}
\definecolor{headertext}{RGB}{255,255,255}

\begin{table*}[htbp]
\centering
\caption{Significant Spearman correlations ($|\rho|>0.40$) among the simulated species (time-averaged value at each of the $n=13$ nodes; bootstrap 95\% intervals in \texttt{table\_spearman.csv}). \colorbox{verystrongbg}{\strut green} = very strong ($|\rho|\ge0.90$), \colorbox{moderatebg}{\strut blue} = moderate ($0.40\le|\rho|<0.60$).}
\label{tab:spearman_corr}
\setlength{\tabcolsep}{8pt}
\renewcommand{\arraystretch}{1.3}
\begin{tabularx}{\textwidth}{>{\raggedright\arraybackslash}p{5.0cm}>{\centering\arraybackslash}p{2.0cm}>{\raggedright\arraybackslash}X}
\toprule
\rowcolor{headerblue}
\textbf{\textcolor{headertext}{Variable pair}} & \textbf{\textcolor{headertext}{$\rho$}} & \textbf{\textcolor{headertext}{Interpretation}} \\
\midrule
\rowcolor{verystrongbg}
\texttt{Dissolved\_Oxygen} $\leftrightarrow$ \texttt{THM} & $-0.985$ & Opposite variation across nodes. \\
\rowcolor{verystrongbg}
\texttt{Conductivity} $\leftrightarrow$ \texttt{THM} & $-0.983$ & Opposite variation across nodes. \\
\rowcolor{verystrongbg}
\texttt{Dissolved\_Oxygen} $\leftrightarrow$ \texttt{Conductivity} & $+0.979$ & Joint variation across nodes; no causal link implied. \\
\rowcolor{moderatebg}
\texttt{Turbidity} $\leftrightarrow$ \texttt{Conductivity} & $+0.456$ & Moderate joint variation. \\
\rowcolor{moderatebg}
\texttt{Chlorine} $\leftrightarrow$ \texttt{THM} & $-0.437$ & Expected sign (chlorine consumed, by-products formed), but only moderate. \\
\rowcolor{moderatebg}
\texttt{Chlorine} $\leftrightarrow$ \texttt{Conductivity} & $+0.409$ & Moderate joint variation. \\
\bottomrule
\end{tabularx}
\end{table*}

\begin{figure}[H]
\centering
\includegraphics[width=\columnwidth]{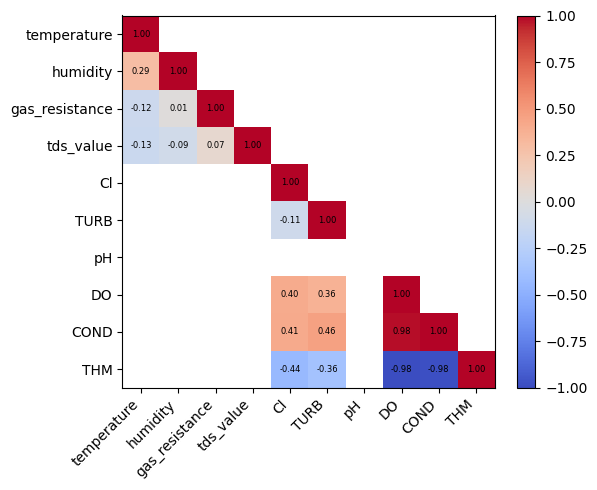}
\caption{Spearman correlation heatmap (\emph{coolwarm} palette, $[-1,1]$ scale) for the main physicochemical parameters: four sensor channels (upper-left block, $n=18{,}787$ real rows without attack or fault) and the simulated species (lower-right block, $n=13$ nodes, time-averaged values). Real--simulated cells are masked and \texttt{pH} is empty (constant across nodes). Only the lower triangle is displayed to avoid redundancy; cell values are rounded to two decimal places.}
\label{fig:corr_matrix}
\end{figure}

Figure~\ref{fig:corr_matrix} presents the lower-triangular heatmap of the Spearman correlation matrix rendered with the \emph{coolwarm} diverging palette: warm red tones indicate strong positive associations, cool blue tones indicate negative ones, and white cells near zero denote the absence of monotone dependence.

All six associations belong to the simulated block. With $n=13$ nodes they describe the spatial structure of the network and not independent observations; the three strongest involve dissolved oxygen, conductivity and THM ($|\rho|\ge0.98$), which vary together across nodes. The expected negative relation between chlorine and THM holds in sign but is moderate: Spearman $\rho(\mathrm{Cl},\mathrm{THM})=-0.44$ (95\% bootstrap CI $[-0.97,\,0.17]$, $n=13$ nodes). The interval includes zero, so the relation is not established at this sample size. The sensor channels show only weak monotone dependence ($|\rho|\le0.29$).

\subsection{Effect of Attacks and Faults on the Sensor Channels}

Because the water-quality label is constant (Section~\ref{sec:labels}), class-conditional distributions carry no information. We instead measure how fault and attack injection modifies the sensor channels of the real rows. At least one of the four sensor channels differs from the raw record in 3{,}568 samples (15.1\%); rows without attack or fault are identical to the raw record. White noise (2{,}368 samples) alters all four channels at once, replay (674 samples) and drift (439 samples) alter the sensor channels listed in Table~\ref{tab:attackchannels}, and 67 spike and 20 extreme samples hit a sensor channel (the other 51 spikes and 27 extreme values act on the Cl/TURB channels). Table~\ref{tab:modified} gives the breakdown per channel.

\begin{table}[H]
\centering
\caption{Real samples whose sensor channel differs from the raw record.}
\label{tab:modified}
\footnotesize
\setlength{\tabcolsep}{3pt}
\resizebox{\columnwidth}{!}{%
\begin{tabular}{lrrrrr}
\toprule
\textbf{Channel} & \textbf{Samples (\%)} & \textbf{White noise} & \textbf{Replay/drift} & \textbf{Spike/extreme} & \textbf{Max.\ deviation} \\
\midrule
Temperature ($^\circ$C) & 2{,}482 (10.5) & 2{,}367 & 85 & 30 & 76.4 \\
Humidity (\%) & 2{,}725 (11.5) & 2{,}368 & 340 & 17 & 59.7 \\
Gas resistance ($\Omega$) & 2{,}549 (10.8) & 2{,}367 & 159 & 23 & 483{,}643 \\
TDS & 2{,}914 (12.3) & 2{,}368 & 529 & 17 & 999.9 \\
\bottomrule
\end{tabular}}
\end{table}

The largest deviations are full-scale faults (for example 483{,}643\,$\Omega$ for gas resistance and 999.9 for TDS).

\section{Machine Learning Model Benchmark}\label{sec:benchmark}

This section first reports the experiment run on the first release of the dataset, then shows why its scores cannot be read as evidence of dataset quality, and finally defines and runs the leakage-aware benchmark. Code for all tasks is in the repository (\texttt{benchmark/}).

\subsection{\texorpdfstring{{Audit of the First Release:} Experimental Protocol}{Audit of the First Release: Experimental Protocol}}

{The experiment below was run on the first release of the dataset, which contained twenty augmented copies of the sensor channels; the released version no longer includes them.}

The benchmarking of classification models is conducted under a standardized experimental protocol.

The adopted protocol is defined as follows:

\begin{itemize}
    \item \textbf{Data splitting:} The dataset (47{,}376 rows) is divided into 80\% training and 20\% testing using stratified \add{random} sampling based on the target variable \textit{quality} to preserve class distribution.
    \item \textbf{Data preprocessing:}
    \begin{itemize}
        \item Categorical variables are encoded using Label Encoding.
        \item {The \texttt{qualite} column is empty on 23{,}688 of 47{,}376 rows; these rows are completed from \texttt{quality\_final} (the two columns agree on every row where both exist). No other imputation is needed.}
        \item Numerical features are standardized using Z-score normalization (StandardScaler), fitted only on the training set to avoid data leakage.
    \end{itemize}
    \item \textbf{Cross-validation:} Stratified 10-fold cross-validation (StratifiedKFold, k = 10) is applied on the training set to ensure model stability and support hyperparameter tuning when required.
    \item \textbf{Evaluation metrics:}
    \begin{itemize}
        \item Accuracy
        \item Macro F1-score (primary metric)
        \item Precision
        \item Recall
    \end{itemize}
    \item \textbf{Reproducibility:} All models use a fixed random seed ($random\_state = 42$).
\end{itemize}

\subsection{Evaluated Models}

Six machine learning algorithms from different families were implemented and evaluated under identical conditions:

\begin{itemize}
    \item \textbf{Random Forest (RF):} Ensemble of 200 decision trees using the Gini impurity criterion.
    \item \textbf{XGBoost:} Gradient boosting model with L1/L2 regularization and early stopping.
    \item \textbf{Support Vector Machine (SVM):} RBF kernel with optimized C and gamma parameters.
    \item \textbf{Multilayer Perceptron (MLP):} Neural network with two hidden layers of 128 neurons, ReLU activation, and Adam optimizer.
    \item \textbf{Logistic Regression:} Multinomial linear model with L2 regularization.
    \item \textbf{K-Nearest Neighbors (KNN):} Classification using k = 5 with Euclidean distance.
\end{itemize}

\subsection{Comparative Results}

The performance results obtained on the test set are summarized in Table~\ref{tab:ml_benchmark}.
\begin{table}[H]
\centering
\footnotesize
\caption{Performance comparison of machine learning models {on the first release (47{,}376 rows) under the v1 protocol (random split, legacy label, all 28 input columns). These scores measure label reconstruction, not generalisation.}}
\label{tab:ml_benchmark}
\renewcommand{\arraystretch}{1.1}
\resizebox{\columnwidth}{!}{%
\begin{tabular}{lcccc}
\hline
\textbf{Model} & \textbf{Accuracy} & \textbf{F1-score (macro)} & \textbf{Precision} & \textbf{Recall} \\
\hline
Random Forest          & \textbf{0.9966} & \textbf{0.9939} & \textbf{0.9962} & \textbf{0.9916} \\
XGBoost                & 0.9949 & 0.9910 & 0.9921 & 0.9900 \\
K-NN                   & 0.9878 & 0.9831 & 0.9834 & 0.9827 \\
MLP                    & 0.9869 & 0.9757 & 0.9821 & 0.9696 \\
SVM (RBF)              & 0.9702 & 0.9576 & 0.9575 & 0.9579 \\
Logistic Regression    & 0.8562 & 0.7991 & 0.8294 & 0.7813 \\
\hline
\end{tabular}%
}
\end{table}

\subsection{Analysis of Results}

\noindent\textbf{Why these scores are not evidence of dataset quality.} Four sources of leakage make near-perfect scores inevitable under the v1 protocol:
\begin{enumerate}
\item \emph{Label is a function of the inputs.} The legacy label is a threshold rule over eight columns (\texttt{tds\_value}, \texttt{temperature}, \texttt{gas\_resistance}, \texttt{humidity}, \texttt{Cl\_J7}, \texttt{Cl\_J3}, \texttt{TURB\_J3}, \texttt{TURB\_J7}) that are themselves inputs; a decision tree trained on these eight columns alone recovers the label with 0.9947 accuracy. Axis-aligned trees recover threshold rules almost exactly (RF and XGBoost both above 0.99 macro-F1), whereas logistic regression, which cannot represent step functions, falls to 0.80. The ranking therefore reflects each model's ability to express thresholds, not ``the non-linear nature of the problem''.
\item \emph{Near-duplicate features.} Twenty of the 28 inputs are jittered, scaled, warped, sliced and permuted copies of four of the label-defining columns (\texttt{tds\_value}, \texttt{temperature}, \texttt{humidity}, \texttt{gas\_resistance}).
\item \emph{Temporal leakage.} At 1--2\,s sampling, consecutive samples are almost identical; a random split places neighbours of each test sample in the training set, so KNN with $k=5$ effectively looks up the answer~\cite{Bergmeir2012CV} (0.983 macro-F1).
\item \emph{Label as feature.} \texttt{label\_attack} was among the inputs of the original run (Table~\ref{tab:shap_global}, rank 8). The fused table contains no duplicated rows (0 of 47{,}376), so row duplication across splits is not a source of leakage here.
\end{enumerate}
Table~\ref{tab:leakage_ablation} removes these sources one at a time. Removing the 20 augmented columns barely changes the score (0.9961 to 0.9870), because the eight label-defining columns remain. Removing those eight as well leaves no sensor input: the 28 inputs are exactly the eight label-defining columns and their 20 augmented copies, and the remaining numeric columns of the table (an attack identifier and a prediction-confidence score) are outputs rather than measurements, so they are not admissible inputs. The last two rows are therefore undefined for the legacy label, which cannot be predicted from any column that is not itself part of its definition. This is the reason the leakage-aware benchmark below moves to tasks whose labels are not derived from the inputs. On the current release (real rows only, eight columns), the same Random Forest reaches 0.976 macro-F1 under a random split but 0.341 under a chronological split, which shows that most of the apparent performance does not survive a chronological split.

\begin{table}[H]
\centering
\caption{Leakage ablation on the first release (47{,}376 rows), Random Forest, macro-F1 (mean $\pm$ std, 5 seeds). ``--'': no input column remains, so the score would only reflect the majority class. The last two rows use the current release (23{,}688 real rows, eight observed channels as inputs, legacy label; random split: 5 seeds; chronological split: single run, seed 42).}
\label{tab:leakage_ablation}
\resizebox{\columnwidth}{!}{%
\begin{tabular}{lc}
\toprule
\textbf{Protocol} & \textbf{Macro-F1} \\
\midrule
v1: random split, all 28 columns, legacy label & 0.9961 $\pm$ 0.0005 \\
\quad $-$ duplicated rows removed (0 found) & 0.9961 $\pm$ 0.0005 \\
\quad $-$ 20 augmented columns removed & 0.9870 $\pm$ 0.0007 \\
\quad $-$ 8 label-defining columns removed & -- \\
\quad random $\rightarrow$ chronological split & -- \\
\midrule
Current release, real rows, 8 columns, random split & $0.9756 \pm 0.0022$ \\
Current release, real rows, 8 columns, chronological split & $0.341$ \\
\bottomrule
\end{tabular}}
\end{table}

\subsection{Global Interpretation}

\subsection{Leakage-Aware Benchmark: Tasks}\label{sec:tasks}
Table~\ref{tab:tasks} defines two tasks, T3 and T4. Neither asks a model to recompute a label from the columns that define it.

\begin{table*}[t]
\centering
\footnotesize
\caption{Benchmark tasks, inputs, targets, splits and metrics.}
\label{tab:tasks}
\renewcommand{\arraystretch}{1.2}
\begin{tabular}{p{3.2cm}p{3.8cm}p{2.6cm}p{3.4cm}p{3.2cm}}
\toprule
\textbf{Task} & \textbf{Input} & \textbf{Target} & \textbf{Split(s)} & \textbf{Metrics} \\
\midrule
T3 Attack and fault classification & window of the eight observed channels (\texttt{temperature}, \texttt{humidity}, \texttt{gas\_resistance}, \texttt{tds\_value}, \texttt{Cl\_J7}, \texttt{Cl\_J3}, \texttt{TURB\_J3}, \texttt{TURB\_J7}) & \texttt{attack\_type} (DoS, FDI, FDI-stealth, drift, replay), \texttt{fault\_type} (extreme, spike, white noise; fault windows are labelled normal in the attack target) & chronological for classification (only replay is present in the test split: FDI occurs in the training split only, stealthy FDI in the validation split only); LOAO by pooling the windows of the held-out type (see text) & macro-F1, per-class recall, detection delay, FAR, LOAO AUROC \\
T4 Synthetic-data utility & synthetic (TimeGAN) windows (the TimeGAN rows carry no attack label) or augmented training data & as T3 & train synthetic / test real (TSTR) & $\Delta$ macro-F1 vs.\ real-only \\
\bottomrule
\end{tabular}
\end{table*}

\noindent\textbf{Split protocols.} \emph{Chronological:} the split is computed on the 23{,}688 real rows only (synthetic rows excluded), ordered in time from 2025-05-19 12:39:08 to 2025-05-20 14:01:00 (nominal step 1~s). The first 60\% of samples (14{,}212, up to 2025-05-20 00:13:55) is used for training, the next 20\% (4{,}737, up to 2025-05-20 12:03:15) for validation (reserved for hyper-parameter selection and early stopping; not used in the runs reported here), and the last 20\% (4{,}739) for testing. Windows ($W{=}30$, stride 5) are built within each split, so that no window overlaps two splits; for T3 this yields 2{,}837 training, 942 validation and 942 test windows. \emph{LOAO} (leave-one-attack-type-out) removes one attack type from training and tests whether a detector trained on the others flags it as anomalous (binary detection); this measures generalisation to unseen threats, the practically relevant case. Because the test split contains only replay, the windows of the held-out type are pooled from the three splits (they are never used for training) and scored against the normal windows of the validation and test splits, with a binary normal/attack target; the AUROC is computed per held-out type and averaged over the five types. Positive and negative windows therefore come from different periods, so a temporal shift can contribute to the score. Official indices are in \texttt{splits.json}.

\noindent\textbf{Models.} Unsupervised (fitted on the normal training windows): Isolation Forest~\cite{Liu2008IForest} (200 trees, contamination 0.05) and one-class SVM (RBF kernel, $\nu=0.05$). Supervised on window statistics (mean, std, min, max, slope, last value and fraction of missing values per channel): RF (200 trees, balanced class weights), XGBoost (200 trees, depth 4) and logistic regression (balanced class weights). Deep models (1D-CNN, LSTM, Transformer, LSTM autoencoder) were not evaluated in this release. All models use default hyper-parameters. Results are mean $\pm$ std over 5 seeds. No significance test is reported, since the best models output a constant prediction.

\subsection{Leakage-Aware Benchmark: Results}

\begin{table*}[t]
\centering
\footnotesize
\caption{T3 attack and fault classification (chronological split) and unseen-attack detection (LOAO). Mean $\pm$ std over 5 seeds. Macro-F1 is computed over the two classes present in the test set (normal, replay); recalls for FDI-stealth and drift are n/a because neither occurs in the test split. Detection delay is averaged over the runs in which at least one attack segment is detected (number of such runs in parentheses when below 5). Deterministic models (std $=0$) give identical results across seeds. LOAO AUROC is the mean over the five held-out attack types (single run, seed 42; per-type values are given in the text); ``--'': not computed.}
\label{tab:t3results}
\renewcommand{\arraystretch}{1.15}
\begin{tabular}{lcccccc}
\toprule
\textbf{Model} & \textbf{Macro-F1} & \textbf{Recall FDI-stealth} & \textbf{Recall drift} & \textbf{Detection delay (s)} & \textbf{FAR (\%)} & \textbf{LOAO AUROC} \\
\midrule
Constant predictor (normal) & $0.465$ & n/a & n/a & n/a & $0.00$ & $0.500$ \\
Isolation Forest & -- & -- & -- & n/a (0/5) & $2.20 \pm 0.11$ & $0.641$ \\
One-class SVM & -- & -- & -- & n/a (0/5) & $5.98 \pm 0.00$ & -- \\
Logistic regression & $0.417 \pm 0.000$ & n/a & n/a & $55.3 \pm 0.0$ & $19.29 \pm 0.00$ & $0.507$ \\
Random Forest & $0.465 \pm 0.000$ & n/a & n/a & n/a (0/5) & $0.00 \pm 0.00$ & $0.644$ \\
XGBoost & $0.465 \pm 0.000$ & n/a & n/a & n/a (0/5) & $0.00 \pm 0.00$ & -- \\
\bottomrule
\end{tabular}
\end{table*}

\begin{table}[H]
\centering
\caption{T4 synthetic-data utility on T3 (Random Forest, macro-F1; mean $\pm$ std over 5 seeds). The test split contains two classes (normal, replay); a constant ``normal'' predictor scores 0.465.}
\label{tab:t4results}
\resizebox{\columnwidth}{!}{%
\begin{tabular}{lc}
\toprule
\textbf{Training data} & \textbf{RF} \\
\midrule
Constant predictor (normal) & $0.465$ \\
Real only & $0.465 \pm 0.000$ \\
Real + augmentation & $0.465 \pm 0.000$ \\
Real + TimeGAN & $0.465 \pm 0.000$ \\
TimeGAN only (TSTR) & n/a \\
\bottomrule
\end{tabular}}
\end{table}

\noindent\textbf{Findings.} (i)~\emph{Composition.} The chronological split gives 2{,}837 training windows (2{,}673 normal, 78 drift, 48 DoS, 20 FDI, 18 replay), 942 validation windows (758 normal, 67 replay, 54 DoS, 53 FDI-stealth, 10 drift) and 942 test windows (819 normal, 123 replay). (ii)~\emph{Classification.} No model exceeds the constant ``normal'' predictor (macro-F1 0.465). Random Forest and XGBoost reach exactly this value because they label every test window normal (replay recall 0, no alarm, hence no detection delay), and their result does not vary with the seed. Logistic regression falls below it (0.417), with a false-alarm rate of 19.3\% and a mean detection delay of 55.3\,s; Isolation Forest and the one-class SVM raise 2.2\% and 6.0\% false alarms and detect no replay segment. (iii)~\emph{Unseen attack types.} The AUROC (RF / logistic regression / Isolation Forest) is 0.784 / 0.236 / 0.797 for DoS, 0.883 / 0.933 / 0.919 for FDI, 0.455 / 0.355 / 0.349 for FDI-stealth, 0.741 / 0.453 / 0.618 for drift and 0.359 / 0.556 / 0.524 for replay, i.e.\ means of 0.64, 0.51 and 0.64. Detection is above chance for DoS, FDI and drift with RF and Isolation Forest, and at or below chance for FDI-stealth and replay, consistent with their in-range values (Section~\ref{sec:perturb}); the a-priori ratings of Table~\ref{tab:attacks} are therefore only partly confirmed (FDI is detected more easily than rated). The FDI estimate rests on a single segment (20 windows), and the positive windows come from all periods whereas the negatives come from the validation and test splits, so a temporal shift may inflate or deflate these values. (iv)~\emph{Synthetic data.} Augmentation (jittering with $\sigma=0.05$ in z-units and scaling, applied to training windows only) and TimeGAN windows (non-overlapping, the four channels absent from TimeGAN set to the training mean) leave the Random Forest unchanged at 0.465 in every configuration, since its prediction stays constant; a paired test is therefore not applicable. The TimeGAN rows carry no attack label, so they cannot supply realistic attack signal, and TimeGAN-only training (TSTR) cannot be evaluated.

Table~\ref{tab:edge} summarises the qualitative characteristics of the candidate models in terms of size and interpretability. No on-device measurements were performed in this study.

\begin{table}[H]
\centering
\caption{Qualitative characteristics of the candidate models (no on-device measurements).}
\label{tab:edge}
\resizebox{\columnwidth}{!}{%
\begin{tabular}{lcc}
\toprule
\textbf{Model} & \textbf{Model size} & \textbf{Interpretability} \\
\midrule
Random Forest   & Medium (grows with trees/depth) & High (feature importance, SHAP) \\
XGBoost         & Small to medium                 & High (feature importance, SHAP) \\
1D-CNN (TFLite) & Small (after quantisation)      & Low (post-hoc only) \\
\bottomrule
\end{tabular}}
\end{table}

\subsection{Final Model Selection}

Based on the experimental results, {Random Forest} is selected as the {reference} model {for the XAI analysis} for the following reasons:

\begin{itemize}
    \item Macro-F1 equal to the best (0.465, tied with XGBoost and the constant predictor) and the highest mean unseen-attack AUROC among the supervised models evaluated (0.64) {under the leakage-aware protocol (Table~\ref{tab:t3results})}
    \item High interpretability through feature importance and SHAP analysis
    \item Lightweight model family, expected to be compatible with embedded deployment {(Table~\ref{tab:edge})}
\end{itemize}
\section{SHAP Methodology (SHapley Additive exPlanations)}\label{sec:xai}

{This section analyses the first release, which contained twenty augmented copies of the sensor channels.}

SHAP is an explainability framework derived from cooperative game theory {~\cite{Lundberg2017SHAP}}. It assigns to each feature a contribution score reflecting its marginal impact on the final model prediction.

For Random Forest models, the TreeSHAP algorithm {~\cite{Lundberg2020TreeSHAP}} allows exact and efficient computation with complexity:
\[
O(TLD^2)
\]
where:
\begin{itemize}
    \item $T$ = number of trees,
    \item $L$ = number of leaves,
    \item $D$ = maximum tree depth.
\end{itemize}

The SHAP value of feature $i$ is defined as:
\[
\phi_i =
\sum_{S \subseteq F \setminus \{i\}}
\frac{|S|!(|F|-|S|-1)!}{|F|!}
\left[f(S \cup \{i\}) - f(S)\right]
\]
where:
\begin{itemize}
    \item $\phi_i$ is the SHAP value of feature $i$,
    \item $S$ is a subset excluding feature $i$,
    \item $f(S)$ is the model prediction using subset $S$.
\end{itemize}

In this study, SHAP computation produced a tensor of size:
\[
(9\,476,\ 28,\ 3)
\]
corresponding respectively to:
\begin{itemize}
    \item 9,476 test samples (20\% of 47,376 rows),
    \item 28 input variables (the eight label-defining columns and their 20 augmented copies),
    \item 3 quality classes (Good, Poor, Very Good).
\end{itemize}

{The Random Forest (200 trees, \texttt{random\_state}=42) was retrained under the v1 protocol on these 28 inputs and reaches an accuracy of 0.9976 and a macro-F1 of 0.9967 on the test split. The 23,688 synthetic rows have no value for 24 of the 28 inputs; these missing values were imputed with the training-set median before standardisation.}

Global importance was obtained by averaging absolute SHAP values over observations and classes:
\begin{verbatim}
mean_shap = np.mean(np.abs(shap_values), axis=(0,2))
\end{verbatim}

\subsection{\texorpdfstring{Global Feature Importance Analysis {(v1 Protocol)}}{Global Feature Importance Analysis (v1 Protocol)}}

Table~\ref{tab:shap_global} presents the nine most influential variables according to their mean absolute SHAP value.
\begin{table*}[!htbp]
\centering
\caption{Top 9 Variables Ranked by Global SHAP Importance {(v1 protocol, legacy label, first release, 47{,}376 rows)}}
\label{tab:shap_global}
\renewcommand{\arraystretch}{1.3}
\setlength{\tabcolsep}{6pt}
\begin{tabular}{clp{2.2cm}p{9.2cm}}
\toprule
\textbf{Rank} & \textbf{Feature} & \textbf{Mean SHAP} & \textbf{Physical Interpretation {/ revised reading}} \\
\midrule
1 & temperature                & 0.1159 & {Air temperature; label input} \\
2 & tds\_value                 & 0.0850 & {Total dissolved solids reading; label input} \\
3 & temperature\_scaled        & 0.0608 & {Randomly rescaled copy of rank 1} \\
4 & temperature\_jitter        & 0.0276 & {Noisy copy of rank 1} \\
5 & tds\_value\_scaled         & 0.0268 & {Randomly rescaled copy of rank 2} \\
6 & gas\_resistance            & 0.0224 & VOC proxy indicating volatile contaminants in surrounding air {--- label input} \\
7 & humidity                   & 0.0170 & {Air humidity; label input} \\
8 & tds\_value\_jitter         & 0.0159 & {Noisy copy of rank 2} \\
9 & temperature\_permuted      & 0.0065 & {Permuted copy of rank 1} \\
\bottomrule 
\end{tabular}
\end{table*}

The feature \texttt{temperature} appears as the dominant predictor, with a SHAP value 36\% higher than the second-ranked variable (\texttt{tds\_value}). {Both are label-defining columns, and so are the other two raw sensor variables in the table (\texttt{gas\_resistance}, \texttt{humidity}). None of the node-level simulated variables (e.g.\ \texttt{Cl\_J3}, \texttt{TURB\_J7}) appears among the nine most influential features. The model therefore relies on the columns that define the legacy label, which is consistent with label reconstruction rather than the detection of a physical water-quality signal.}

Temperature and \texttt{tds\_value} lead the ranking, and the third variable is itself a rescaled copy of temperature, highlighting the strong role of the label-defining inputs and of their near-duplicates { in the legacy label}.

A marked drop is observed after rank 3 (0.0608 to 0.0276), and ranks 4 and 5 are close to each other (0.0276 and 0.0268). {Of the nine variables, six are augmented copies of \texttt{temperature} or \texttt{tds\_value}, which shows how the duplicated features share the importance of their source column.}

\subsection{Analysis of Engineered Feature Redundancy}

{The augmented columns are not normalised or variability features: \texttt{\_scaled} is the raw value multiplied by a random factor, and \texttt{\_jitter} is the raw value plus random noise. When correlated copies of a feature are present, tree models split importance among them arbitrarily, so the combined share of a source column and its five copies is the relevant quantity. Summing mean $|\phi|$ over each source column and its copies (path-dependent TreeSHAP, 9{,}476 test rows), \texttt{temperature} accounts for 0.216 of the total (50.1\%), \texttt{tds\_value} for 0.142 (33.0\%), \texttt{gas\_resistance} for 0.040 (9.4\%) and \texttt{humidity} for 0.031 (7.2\%); the four columns \texttt{Cl\_J3}, \texttt{TURB\_J3}, \texttt{Cl\_J7} and \texttt{TURB\_J7} together account for 0.001 (0.3\%). The three ambient variables and their copies thus carry about 67\% of the total importance.}

{SHAP attributions are known to be unreliable for strongly correlated inputs ~\cite{Janzing2020,Aas2021}, so we repeated the analysis in three ways. (i)~An interventional TreeSHAP (background of 100 training rows, 1{,}000 explained test rows) gives the same group ordering (\texttt{temperature} 51.8\%, \texttt{tds\_value} 33.9\%) but changes individual ranks: \texttt{temperature\_scaled} (0.0867) moves above \texttt{tds\_value} (0.0850) and just below \texttt{temperature} (0.1046). (ii)~After removing the 20 copies, the Random Forest trained on the eight label-defining columns still reaches an accuracy of 0.9904 and a macro-F1 of 0.9866; interventional importance concentrates on \texttt{temperature} (0.193) and \texttt{tds\_value} (0.123), followed by \texttt{gas\_resistance} (0.031), \texttt{humidity} (0.024) and the four \texttt{Cl}/\texttt{TURB} columns (0.017--0.023). (iii)~On the 23{,}688 real rows only, which avoids the median imputation of the synthetic block, macro-F1 is 0.9998 and seven of the nine most important variables are the same as in the full run; eight of the nine belong to the \texttt{temperature} or \texttt{tds\_value} groups, but the order within the groups changes (\texttt{temperature\_scaled} ranks first and the raw \texttt{temperature} third). The individual ranking of a column and its copies is therefore not stable, whereas the group-level result is. Removing the copies does not make the model valid: the eight remaining columns still define the legacy label.}

\subsection{Design Implications and Practical Insights}

The XAI analysis highlights three major design implications:

\begin{enumerate}
    \item
    {\textbf{Explainability can audit a dataset.} Global SHAP on the v1 protocol showed that the model relies almost only on the label-defining columns and their augmented copies (99.7\% of the total importance; \texttt{Cl}/\texttt{TURB} columns 0.3\%), two leakage paths that accuracy alone had hidden.}
    \item
    {\textbf{Labels must not be functions of the inputs.} A water-quality label should be derived from water-quality parameters and must not share columns with the model inputs; otherwise the classifier only reconstructs the labelling rule. Here, the ambient variables (temperature, humidity, gas resistance) and their copies carry about 67\% of the importance.}
    \item
    {\textbf{Compact detection (future work).} Whether a model restricted to the highest-ranked window features keeps its detection performance on the leakage-aware tasks was not tested here.}
\end{enumerate}

\begin{quote}
\textbf{Methodological Note.} SHAP satisfies key axiomatic properties such as efficiency, symmetry, and consistency. Unlike impurity-based Random Forest importance measures, SHAP does not systematically favour continuous high-cardinality variables {; it does, however, share credit among correlated features, which must be taken into account when interpreting rankings}.
\end{quote}
\section{\texorpdfstring{{Data Availability and Reuse}}{Data Availability and Reuse}}\label{sec:availability}

\noindent\textbf{Repository.} The data tables (v1.0, CC BY 4.0) are archived on Zenodo at \url{https://doi.org/10.5281/zenodo.23390541}. The source code (MIT), the EPANET/EPANET-MSX model files, the data dictionary (\texttt{dictionary.csv}), the datasheet and the official splits (\texttt{splits.json}) are available on GitHub at \url{https://github.com/nejiamal3-oss/aquasense}. Files are provided as CSV.

\noindent\textbf{License.} Data CC BY 4.0; code MIT.

\noindent\textbf{Versioning and extension.} Future acquisitions and simulations will be released as new versions under the same concept DOI with an unchanged schema, so results remain comparable.

\noindent\textbf{Intended and non-intended uses.} Intended: benchmarking anomaly and attack detectors, forecasting, synthetic-data and domain-adaptation research. Not intended: drawing physical conclusions from relations between real and simulated channels, or certifying the potability of real water.

\section{Conclusion and Future Directions}\label{sec:conclusion}

\subsection{Summary of Contributions}

This paper presented AquaSense-ICS, an open dataset that co-registers a real 25.4\,h recording (8.0\,h of recorded data) with a six-species EPANET-MSX network model and adds a seeded library of attacks and faults with sample-level labels. Answering the research questions: (RQ1) the attack and fault labels are traceable to a seeded generator, and the water-quality label to documented thresholds; the water-quality label is constant on the available data; (RQ2) random splits with feature-derived labels inflate macro-F1 to 0.9998 on the first release; on the current release the same classifier falls from 0.976 to 0.341 macro-F1 when the random split is replaced by a chronological one; (RQ3) under chronological and leave-attack-type-out splits, no model exceeds the constant predictor in macro-F1 (0.465), and the best detectors reach a mean AUROC of 0.64 on unseen attack types, with detection above chance for DoS, FDI and drift and at or below chance for replay and stealthy FDI. The benchmark offers a template for evaluating other hybrid real--simulated datasets. 

\subsection{Limitations}

The current version of the dataset presents several limitations that should be acknowledged by users:

\begin{itemize}
    \item The real sensor acquisition phase covers a {short (25.4\,h recording window, 8.0\,h recorded)}, limited geographic deployment involving a single indoor installation, which may not fully capture the spatial heterogeneity of large-scale distribution networks {or seasonal effects}.
    \item The number of physical sensor modalities remains constrained by hardware availability. {Only TDS is measured in the water; temperature, humidity and gas resistance describe the ambient air, and no immersed temperature probe was available for TDS compensation.} Additional real-sensor parameters such as pH, dissolved oxygen, and turbidity would further enrich the dataset.
    \item {The real node and the simulated network are independent systems aligned on a common clock; the dataset does not contain physically coupled real--simulated measurements.}
    \item {Water-quality labels are rule-based and have not been validated against laboratory grab samples.}
    \item {Attacks are injected into recorded data (data-level attacks); they do not model the closed-loop physical consequences of manipulated control actions.}
    \item The simulated network topology, while representative, does not correspond to a specific real-world infrastructure, which may limit direct applicability to site-specific engineering studies.
    \item Only 11 of the 24 simulated hourly states are matched by real samples, so the simulated contamination events are only partly visible in the fused table, and the turbidity pulse at J7 is not matched by any real sample.
    \item The water-quality label is constant on the available data, so it cannot serve as a classification target.
    \item The TDS values are not calibrated and carry no unit.
    \item Three of the five attack types (FDI, stealthy FDI and DoS) act only on simulated channels.
    \item The release contains 14 attack segments (FDI and stealthy FDI one each), so attack types cannot be spread over the training, validation and test splits; generalisation to unseen attack types is evaluated only by pooling windows across splits, from very few segments per type (a single one for FDI and stealthy FDI), and the resulting AUROC values are indicative.
    \item The disturbances applied to the sensor were not recorded in a machine-readable log.
\end{itemize}

\subsection{Future Research Directions}

Several extension and improvement directions for 

 the  AquaSense-ICS dataset have been identified:

\begin{enumerate}
    \item \textbf{Expansion of Real Data Acquisition.} Future versions will extend the real data collection phase to multiple field deployments across geographically diverse sites. Additional physicochemical sensors, such as pH electrodes, dissolved oxygen optical probes, and nephelometric turbidity sensors, will be integrated{, together with an immersed temperature probe and periodic laboratory grab samples for label validation}. This will enrich the real sensing modalities and improve ecological representativeness.

    \item \textbf{Larger and More Complex Simulated Networks.} The simulated network topology will be expanded toward larger and more complex water distribution systems, including multi-zone networks with elevated tanks, booster pumping stations, and pressure management areas{, starting with the public L-TOWN and C-Town models so that results can be compared with BattLeDIM and BATADAL}. This will enable spatially distributed simulation data with finer hydraulic resolution.

    \item \textbf{Enhanced Cyber-Physical Attack Library.} The cyber-physical attack library will be enriched with new categories designed to increase realism, including coordinated multi-point contamination injection scenarios, physically coherent sensor spoofing patterns, and stealthy low-rate intrusions.

    \item \textbf{Extended Temporal Coverage.} The dataset timeline will be expanded to cover full seasonal cycles, capturing the effects of temperature variation, demand seasonality, and environmental factors on sensor readings and hydraulic behavior.

    \item \textbf{Refined Labeling Methodology.} The annotation methodology will be improved through collaboration with water quality domain experts to integrate regulatory thresholds from multiple national standards, resulting in a more nuanced and internationally applicable quality labeling system.

    \item {\textbf{Federated and foundation-model studies:} With multiple deployment sites, the dataset will support federated learning across sites and the evaluation of pre-trained time-series foundation models, which the present single-site version cannot meaningfully test.}
\end{enumerate}


\section*{CRediT authorship contribution statement}
A.N: Conceptualization, Methodology, Software, Investigation, Writing -- original draft.

M.F: Data curation, Validation, Writing -- review \& editing.

L.C.F: Data curation, Validation, Writing -- review, editing \& Supervision.

\section*{Declaration of competing interest}
The authors declare that they have no known competing financial interests or personal relationships that could have appeared to influence the work reported in this paper.

\section*{Data availability}
The data (CC~BY~4.0) are archived on Zenodo at
\url{https://doi.org/10.5281/zenodo.23390541} and the source code (MIT) is
available on GitHub at \url{https://github.com/nejiamal3-oss/aquasense}
(see Section~\ref{sec:availability}).

\section*{Declaration of generative AI and AI-assisted technologies in the manuscript preparation process}
During the preparation of this work the authors used Claude (Anthropic) in order to revise the wording and the LaTeX structure of the manuscript. After using this tool, the authors reviewed and edited the content as needed and take full responsibility for the content of the published article.
\section*{Acknowledgements}
{This research received no specific grant from any funding agency in the public, commercial or not-for-profit sectors. / Funding sources and acknowledgements.}

\bibliographystyle{elsarticle-num}
\bibliography{example_revised}

\end{document}